\documentclass[12pt,twocolumn]{aastex62}
\usepackage[left]{lineno}

\usepackage{hyperref}
\usepackage{amsmath,mathtools,mathrsfs,amssymb}
\usepackage{graphicx}
\usepackage{bm}
\usepackage[percent]{overpic}

\newcommand{\bete}[1]{\textcolor{red}{#1}}

\defcitealias{medinatorrejon_etal_2021}{MGK+21}
\defcitealias{kadowaki_etal_2021}{KGM+21}
\graphicspath{{./}{figures/}}

\accepted{May 16, 2023}
\submitjournal{ApJ}

\shorttitle{Particle acceleration by magnetic reconnection in relativistic jets}
\shortauthors{Medina-Torrejon, de Gouveia Dal Pino and  Kowal}

\usepackage{multirow}

\begin{document}

\title{Particle acceleration by magnetic reconnection in relativistic jets: the transition from small to large scales}

\correspondingauthor{Tania E. Medina-Torrejon
}
\email{temttm@gmail.com}

\author[0000-0003-4666-1843]{Tania E. Medina-Torrej\'{o}n}           
\author[0000-0001-8058-4752]{Elisabete M. de Gouveia Dal Pino}
\affiliation{Universidade de S\~{a}o Paulo, Instituto de Astronomia, Geof\'{i}sica e Ci\^{e}ncias Atmosf\'{e}ricas, Departamento de Astronomia, 1226 Mat\~{a}o Street, S\~{a}o Paulo, 05508-090, Brasil} 

\author[0000-0002-0176-9909]{Grzegorz Kowal}
\affiliation{Escola de Artes, Ci\^encias e Humanidades - Universidade de S\~ao Paulo,
Av. Arlindo B\'ettio, 1000 -- Vila Guaraciaba, CEP: 03828-000, São Paulo - SP, Brazil}




\begin{abstract}

 Several MHD works and, in particular, the recent one by 
 Medina-Torrejon et al. (2021) based on three-dimensional MHD simulations of relativistic jets, have evidenced that particle  acceleration by magnetic reconnection driven by the  turbulence in the flow occurs from the resistive  up to the large injection scale of the turbulence.  Particles experience   Fermi-type acceleration up to ultra-high-energies, predominantly  of the parallel velocity component to the local magnetic field, in the reconnection layers in all scales due to the ideal electric fields of the background fluctuations ($V\times B$, where $V$  and $B$ are the velocity and magnetic field of the  fluctuations, respectively). In this work, we show MHD-particle-in-cell (MHD-PIC) simulations following the early stages of the particle acceleration in the relativistic jet which confirm these previous results, demonstrating the strong potential of magnetic reconnection driven by turbulence to accelerate relativistic particles  to extreme energies in magnetically dominated flows. 
 Our results also show that the dynamical time variations of the background magnetic fields do not influence the acceleration of the particles in this process.

\end{abstract}



\section{Introduction} \label{sec:intro}

The role of magnetic reconnection  in the acceleration of energetic particles has lately gained tremendous importance in high energy astrophysics \citep{dalpino_lazarian_2005,giannios_etal_09,dgdp_etal_10,zhang_yan_11,hoshinoLyu2012,mckinney_2012,arons2013,kadowaki_etal_15, singh_etal_15,zhangli2015,zhang_etal_2018}. 
It is now regarded as a strong candidate for the production
of ultra-high energy cosmic rays (UHECRs)  \citep[e.g.][]{medinatorrejon_etal_2021} and very high energy (VHE) flares in the magnetically dominated regions of relativistic sources (i.e., where the magnetic energy is of the order or exceeds the rest mass energy of the particles) \citep[e.g.,][]{cerutti_etal_2013,yuan_etal_2016,lyutikov_etal_2018,petropoulou_etal_2016,christie_etal_19,mehlhaff_etal_2020,kadowaki_etal_2021}.


The comprehension of particle acceleration driven by magnetic reconnection has greatly improved  thanks to both  particle-in-cell (PIC) simulations (predominantly performed in two-dimensions - 2D) 
\citep[e.g.,][]{zenitani_H_2001,drake_etal_2006,zenitani_H_2007,zenitani_H_2008,lyubarsky_etal_2008,drake_etal_2010,clausen-brown_2012,cerutti_etal_2012,cerutti_etal_2014,li_etal_2015,werner_etal_2018,werner_etal_2019,lyutikov_etal_2017,sironi_spitkovsky_2014,guo_etal_2015,guo_etal_2016,guo_etal_2020,
sironi_etal_2015,
ball_etal_2018,kilian_etal_2020,sironi2022}), 
and MHD  simulations (generally performed in 3D) 
\citep[e.g.,][]{kowal_etal_2011,kowal_etal_2012,delvalle_etal_16,beresnyak_etal_2016,guo_etal_2019, medinatorrejon_etal_2021}.
They both have established reconnection as an efficient process of acceleration.

Our understanding is that particles are predominantly accelerated in reconnection sites by a Fermi-type mechanism in ideal electric fields \citep{dalpino_lazarian_2005,drake_etal_2006,kowal_etal_2012,guo_etal_2019}.
They undergo multiple crossings in the two converging magnetic fluxes of opposite polarity moving to each other at the reconnection velocity ($V_{rec}$), thereby gaining energy from
head-on interactions with background magnetic irregularities \citep[see also][for reviews] {lazarian_etal_2012,dalpino_kowal_15,lazarian20}.
In order to produce fast reconnection and hence, efficient particle acceleration, the  ubiquitous turbulence in astrophysical MHD flows is 
 acknowledged as one of the main driving mechanisms.   The wandering of the magnetic field lines in the turbulent flow allows for many simultaneous events of reconnection and the enlargement of the outflow regions, removing the reconnected flux more efficiently. These two factors result in the reconnection rate being a substantial fraction of the Alfvén speed and  independent of the microscopic magnetic resistivity  (i.e., independent of the Lundquist number and depending only on the parameters of the turbulence) \citep[][]{lazarian_vishiniac_99,kowal_etal_09,eyink2013,takamoto_etal_15,santoslima_etal_2010,santos-lima_etal20,lazarian20}. 
 The intrinsic 3D nature of  the turbulent reconnection 
 and the particle acceleration that it entails makes the process more efficient than the acceleration in the 2D shrinking plasmoids and X-points that are usually excited by tearing mode instability in  PIC \citep{hoshinoLyu2012,drake_etal_2006,sironi_spitkovsky_2014} and in resistive MHD  \citep[e.g.,][]{kowal_etal_2011,puzzoni_etal_2022} simulations. Moreover, 2D plasmoids are nothing but the cross section of 3D reconnecting magnetic flux tubes, and particle acceleration in nature cannot be confined to 2D plasmoids.
 This has been successfully verified in  3D MHD simulations considering  the injection of thousands of test particles in a current sheet with embedded forced turbulence 
 \citep{kowal_etal_2012,delvalle_etal_16}. In these simulations, the formation of a thick volume filled with large number of converging  reconnecting layers covering the entire inertial range of the turbulence, from the resistive to the injection scale, allows particle acceleration up to the very large scales of the system and to high energies. 
These are crucial  differences with regard to PIC simulations  which can probe only the kinetic small (resistive) scales of the acceleration process, dealing  with  large intrinsic  resistivity wherein particles are predominantly accelerated by  non-ideal electric fields and only up to a few thousand times their rest mass energy. 
Due to these differences one has to be very cautious
when extrapolating the results of particle acceleration from PIC simulations to the macroscopic scales of real systems \citep[see e.g. review in][]{lazarian_etal_2012}. 

The MHD studies mentioned above \citep{kowal_etal_2012,delvalle_etal_16}  considered particle acceleration in non-relativistic domains of 3D reconnection. 
More recently, \citet{medinatorrejon_etal_2021} \citepalias[hereafter][]{medinatorrejon_etal_2021} and \citet{kadowaki_etal_2021} \citepalias[hereafter][]{kadowaki_etal_2021}, motivated by current debates related to the origin of cosmic ray acceleration and VHE variable emission in relativistic jets, and specially in blazars \citep[e.g.,][]{aharonian_etal_07,ackermann_etal_2016,britto_elal_2016,aartsen_etal_2018},
investigated particle acceleration 
in a 3D relativistic magnetically dominated  jet 
subject to  current driven kink  instability (CDKI), by means of relativistic MHD simulations \citep[using the \texttt{RAISHIN} code;][]{mizuno_etal_12, singh_etal_16}.
The instability drives  turbulence and fast magnetic reconnection in the jet flow. 
Its  growth and saturation causes the excitation of large amplitude wiggles along the jet and the disruption of the initial helical magnetic field configuration, leading to the
 formation of several sites of fast reconnection. The turbulence developed  follows approximately a Kolmogorov spectrum 
 \citepalias[][]{kadowaki_etal_2021}.
Test protons injected in the nearly stationary snapshots of the jet,  experience an exponential acceleration in time, predominantly its momentum component  parallel to the local field,
up to a maximum energy.  For a background magnetic field of $B \sim 0.1$ G, this saturation  energy is 
$\sim 10^{16}$ eV, while for $B \sim 10$ G it is  $\sim 10^{18}$ eV. 
There is a clear association of the accelerated particles with the regions of fast reconnection and largest current density.
The particles interact with magnetic fluctuations from the small dissipative 
scales up to the injection scales of the turbulence, which is of the order of the size of the jet diameter. For this reason, the Larmor radius of the particles attaining the saturation energy, which gives the maximum size of the acceleration region, is also of the same order.
 Beyond the saturation value, the particles suffer further acceleration to energies up to 100 times larger, but at a slower rate, due to drift in the   largest scale non-reconnecting fields.
The energy  spectrum of the accelerated particles develops a  high energy tail with a power law index $p \sim$ -1.2 in the beginning of the acceleration, in agreement with earlier works \citepalias[][]{medinatorrejon_etal_2021}. 

In this work, we present results of 3D MHD-PIC simulations of relativistic jets \citep[using the \texttt{PLUTO} code;][]{mignone_etal_2018}, considering in most of the tests the same initial jet setup as in 
\citetalias[][]{medinatorrejon_etal_2021} and 
\citetalias[][]{kadowaki_etal_2021}. Our main goals here are: (i) to test the early stages of the acceleration of the particles evolving at the same time that the jet develops the turbulence driven by the CDKI; (ii) to compare with these previous studies which were performed with test particles launched in the MHD jet after it achieved a nearly steady state regime of fully developed  turbulence; and (iii) to investigate potential effects of the background magnetic field dynamical time evolution on particle acceleration. We find that the results are very similar to the previous studies.  Particles are accelerated by the ideal electric field of the background fluctuations in the reconnection layers of the turbulent flow, from the small resistive scale up to the large injection scales of the turbulence. 
Furthermore, the time evolution of the background fields does not affect their acceleration.




The paper is organized as follows, in Section 2 we describe the numerical method and setup, in Section 3, the results we obtained from the numerical simulations, and in Section 4 we discuss the results and draw our conclusions. 

\section{Numerical Method and Setup} \label{sec:numethod}

We performed 3D relativistic MHD-PIC  simulations of a jet using 
the \texttt{PLUTO} code with non explicit resistivity \citep{mignone_etal_2018}. 
We  employed the HLLD Riemann solver  to calculate the fluxes \citep*{mignone_u_b_2009}, a flux-interpolated constrained  transport to control the divergence $\nabla \cdot B = 0$ \citep{mignone_etal_2019}, and a second-order TVD Runge–Kutta scheme to advance the equations in time. 

We have used a similar setup as in \citetalias{medinatorrejon_etal_2021} and \citetalias{kadowaki_etal_2021},  considering a rotating relativistic jet with initial force-free helical magnetic field and  initial decreasing radial density profile \citepalias[for more details, see][]{medinatorrejon_etal_2021} and also \citep{mizuno_etal_12,singh_etal_16}. 

The computational domain in  Cartesian coordinates $(x, y, z)$ has dimensions $10L \times 10L \times 6L$,  where $L$ is the length scale unit. The larger domain adopted in the x and y directions is  due to the fact that the jet structure exceeds the boundaries of the box in evolved times.  We have imposed outflow boundaries in the transverse directions x and y and  periodic boundaries in the z direction.  We have considered in most of the simulations  a   grid resolution with $256$ cells in each direction, (implying a cell size in the z direction of $\sim$0.02 L, and in the x and y directions of 0.04 L), but in order to test the convergence of the results we have also run a 
model with  $426$ cells in the x and y directions and 256 in e the z direction, (implying a cell size of $\sim$0.02 L in all directions). 


The code unit (c.u.) for the velocity is the light speed  $c$, for time is $L/c$,  for  density  is $\rho_0 = $1, for  magnetic field 
is  $\sqrt{4 \pi \rho_0 c^2}$,  and for  pressure is  $\rho_0 c^2$. 

We have considered two different initial values of the magnetization parameter  $\sigma_0 = B_0^2/\gamma^2 \rho h \sim 0.6$  and $10$ at the jet axis, corresponding to a magnetic field $B_0 = 0.7$ and density $\rho = 0.8$, and $B_0= 4.0$ and $\rho= 1.6$, respectively, where $\gamma$ is the Lorentz factor and $h$ is the specific enthalpy (with $\gamma \sim 1$ and $h \sim 1$ at the axis). Hereafter, We will refer to these models simply as the $\sigma \sim 1$ and $\sigma \sim 10$ models.



In order to drive turbulence in the  jet, we allow for the development of the current-driven-kink instability (CDKI) by  imposing an initial perturbation in the radial velocity profile as in \citetalias{medinatorrejon_etal_2021} \citep[equation 7; see also][]{mizuno_etal_12,singh_etal_16}.

In the MHD-PIC mode, the test particle trajectories are integrated in the time evolving plasma fields (velocity and magnetic ) using
the Boris pusher method \citep{boris1970} which requires the definition of the 
 charge-to-mass ratio for the particles. We have adopted here  $e/mc = $ 20,000,  which implies a physical length scale relation in cgs units: 
 \begin{equation}
  \left ( \frac{e}{mc} \right ) = \left ( \frac{e}{mc} \right )_{cgs} L_{cgs} \sqrt{\rho_{cgs}}
 \end{equation}
Where $e$ and $m$ are the particle charge and mass, respectively. We have adopted $\rho_{cgs} = 1.67 \times 10^{-24} g$ $cm^{-3}$ (or  $n_{cgs} =1$  $cm^{-3}$), which results  a physical length scale  $L_{cgs} \sim 5.2 \times 10^{-7}$ pc. In most of the models, we integrated the trajectories of 10,000 - 50,000 protons with initial uniform space distribution inside the domain, and initial kinetic energies between $(\gamma_p -1) \sim$  1 and 200, where $\gamma_p$ is the particle Lorentz factor, with velocities randomly generated by a Gaussian distribution.


Besides employing the MHD-PIC mode of the \texttt{PLUTO} code to investigate particle acceleration, we have also considered a model where we injected  test particles after the full development of  turbulence  in the jet flow, as in \citetalias{medinatorrejon_etal_2021}. This test was performed with the \texttt{GACCEL} code \citep{kowal_etal_2012,medinatorrejon_etal_2021}. 

We further notice  that, in order to make direct comparisons of the MHD-PIC simulations with the previous work involving test particle injections in frozen-in-time MHD fields, 
we did not account for the accelerated particles feedback on the background plasma, which will be considered in forthcoming work.

~\\
\section{Results}\label{sec:results}

\begin{figure*}[h] 
\centering
   \includegraphics[scale=0.3]{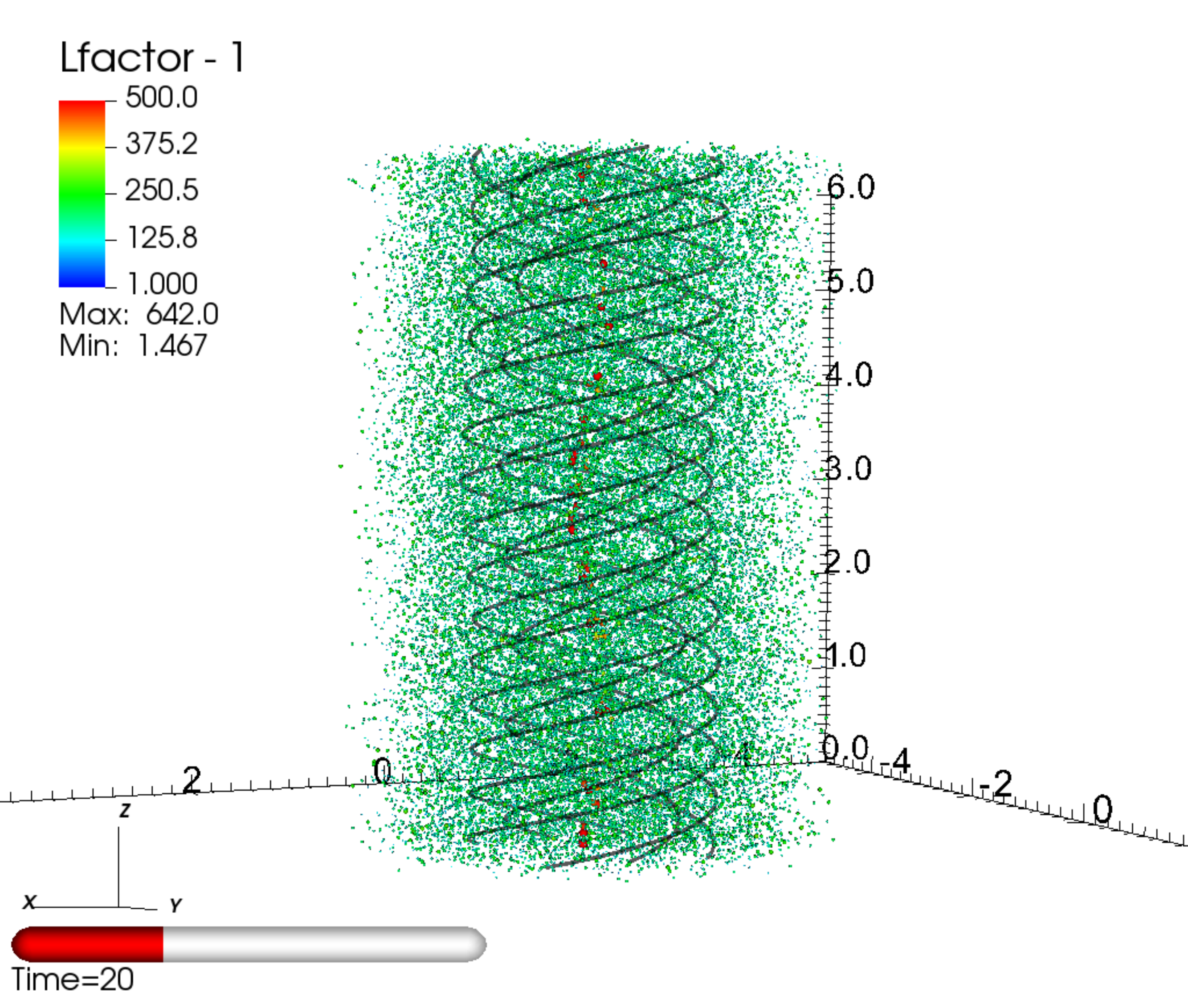}
   \includegraphics[scale=0.3]{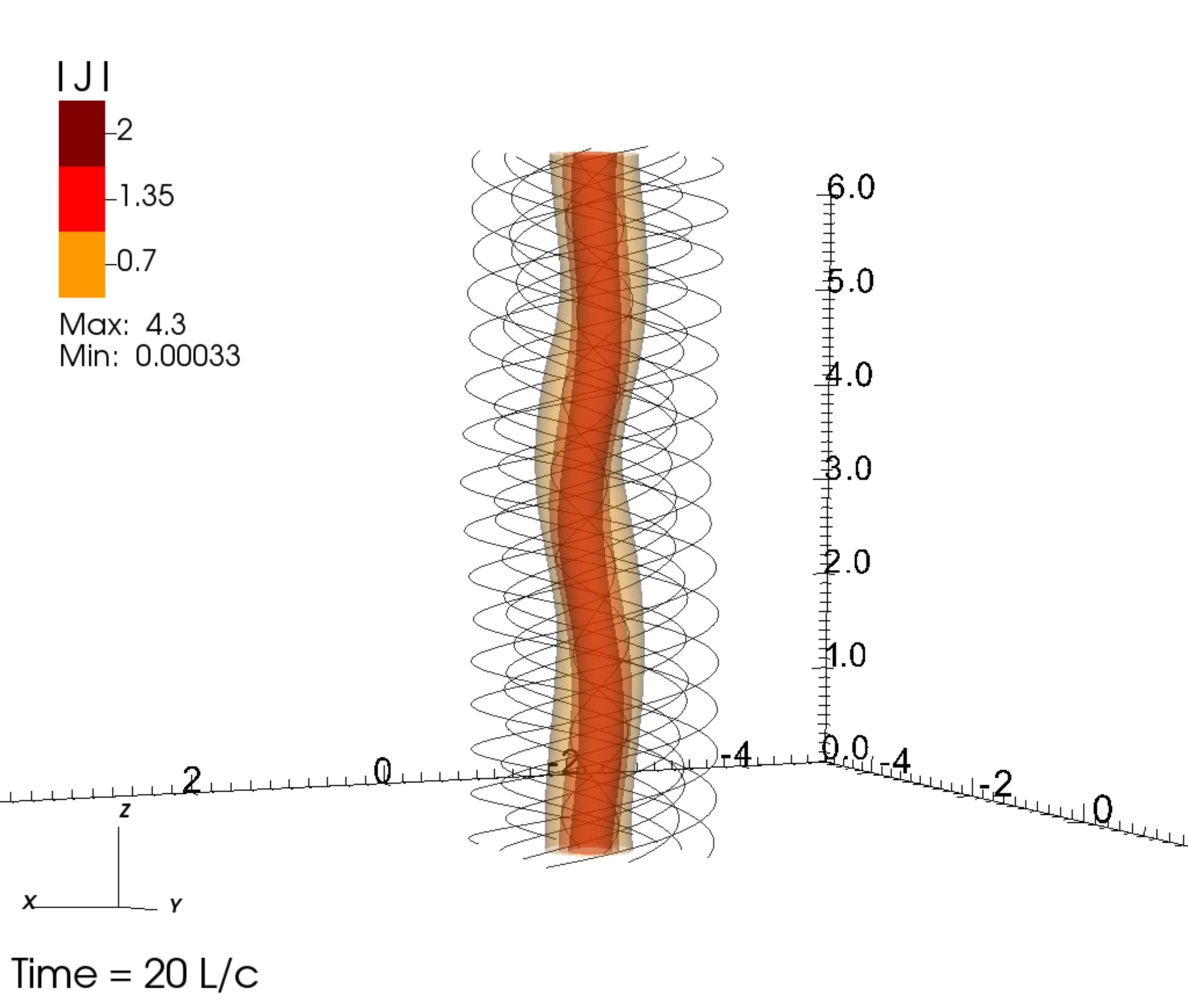}
   \includegraphics[scale=0.3]{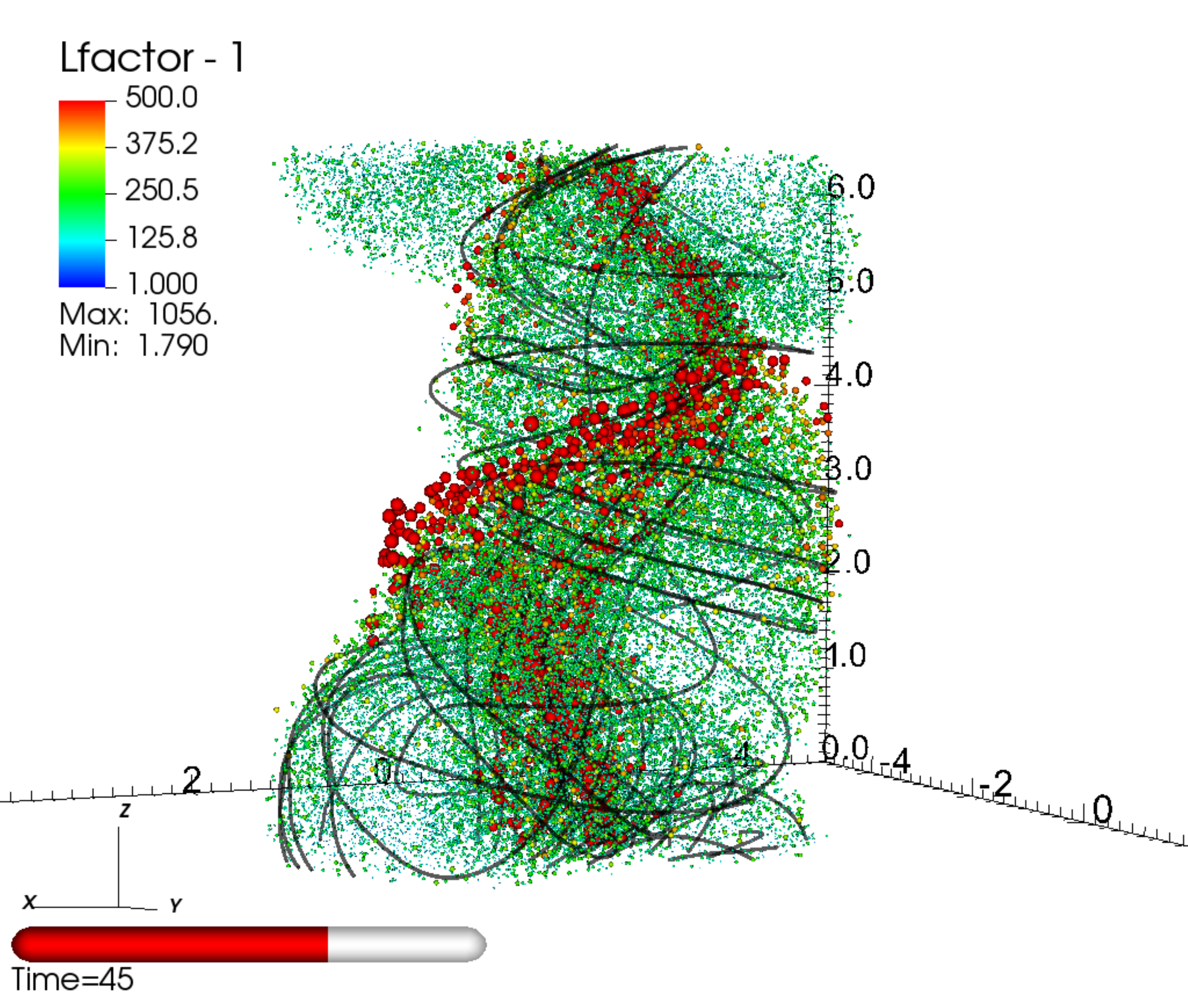}
   \includegraphics[scale=0.3]{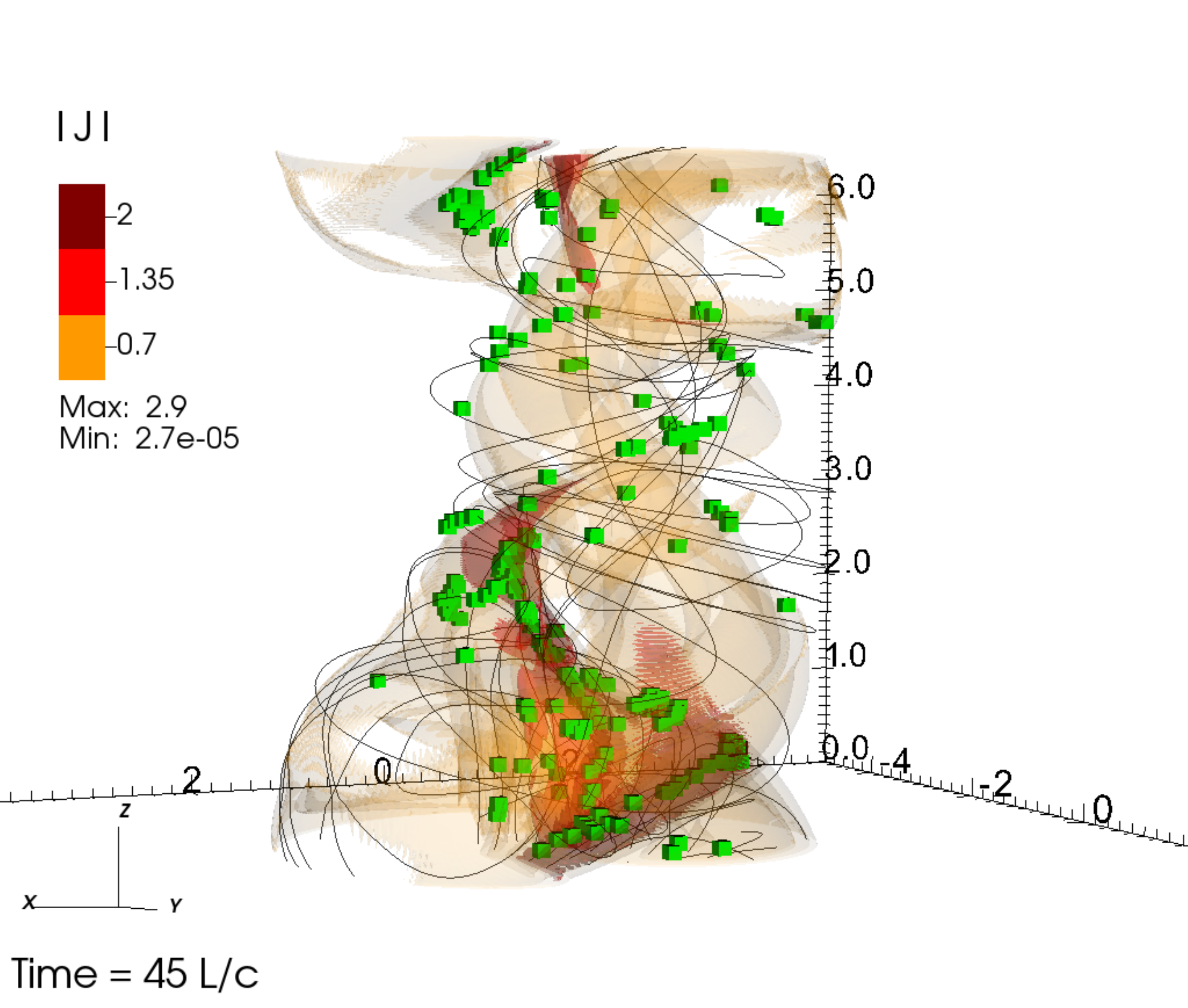}
\caption{Three dimensional view of the  $\sigma \sim 1$ jet evolved with the MHD-PIC mode at t = 20  (top), and 45 L/c (bottom). Left panels: the black lines represent the magnetic field, and the circles the 50,000 particles distribution. The color and size of the circles indicate the value of their kinetic energy normalized by the rest mass energy ($\gamma_p -1$). Right panels: the orange color represents  iso-surfaces of half of the maximum of the current density intensity $|J|$, the black lines the magnetic field, and the green squares  correspond to the positions of the fastest magnetic reconnection events,  with reconnection rate $\geq 0.05$. See text for more details.}
\label{jet_points}
\end{figure*}

Figure \ref{jet_points} shows  the  $\sigma \sim 1$ jet evolved with the MHD-PIC mode of the PLUTO code (with a resolution $256^3$) for two snapshots. A total of 50,000 particles were initially injected in the system. The dynamical  evolution of the jet is very similar to the one obtained in \citetalias{medinatorrejon_etal_2021} and \citetalias{kadowaki_etal_2021} with the \texttt{RAISHIN} MHD code.  With the growth of  the CDKI, the initial helical magnetic field structure 
 starts to wiggle (see $t=20$ L/c) and then,  turbulence develops distorting entirely the field lines and driving fast magnetic reconnection sites, as we see in the right panel for $t=45 L/c$. We note that there are already a few particles being accelerated in the wiggling jet spine at $t=20 L/c$ (left top panel). This is due to curvature drift acceleration, as detected also in the PIC simulations by \citet{alves_etal_2018}, and  in \citetalias{medinatorrejon_etal_2021} with  test particles injected in a similar snapshot of the  background MHD jet (see their Figure 6). Nevertheless, massive particle acceleration takes place only later on, when turbulence and fast reconnection fully develops in the system, as indicated in the left bottom panel at $t=45 L/c$.  The correlation of the accelerated particles (represented by the red   circles with increasing diameter as the energy increases) 
 with the sites of high current density and fast reconnection  (right bottom panel) is evident. A very similar result was obtained for the $\sigma \sim$ 1 jet model run with  larger resolution  ($426^2 \times 256$). In the next paragraphs, we will further quantify these associations.

\begin{figure} %
  \centering
 \includegraphics[scale=0.6]{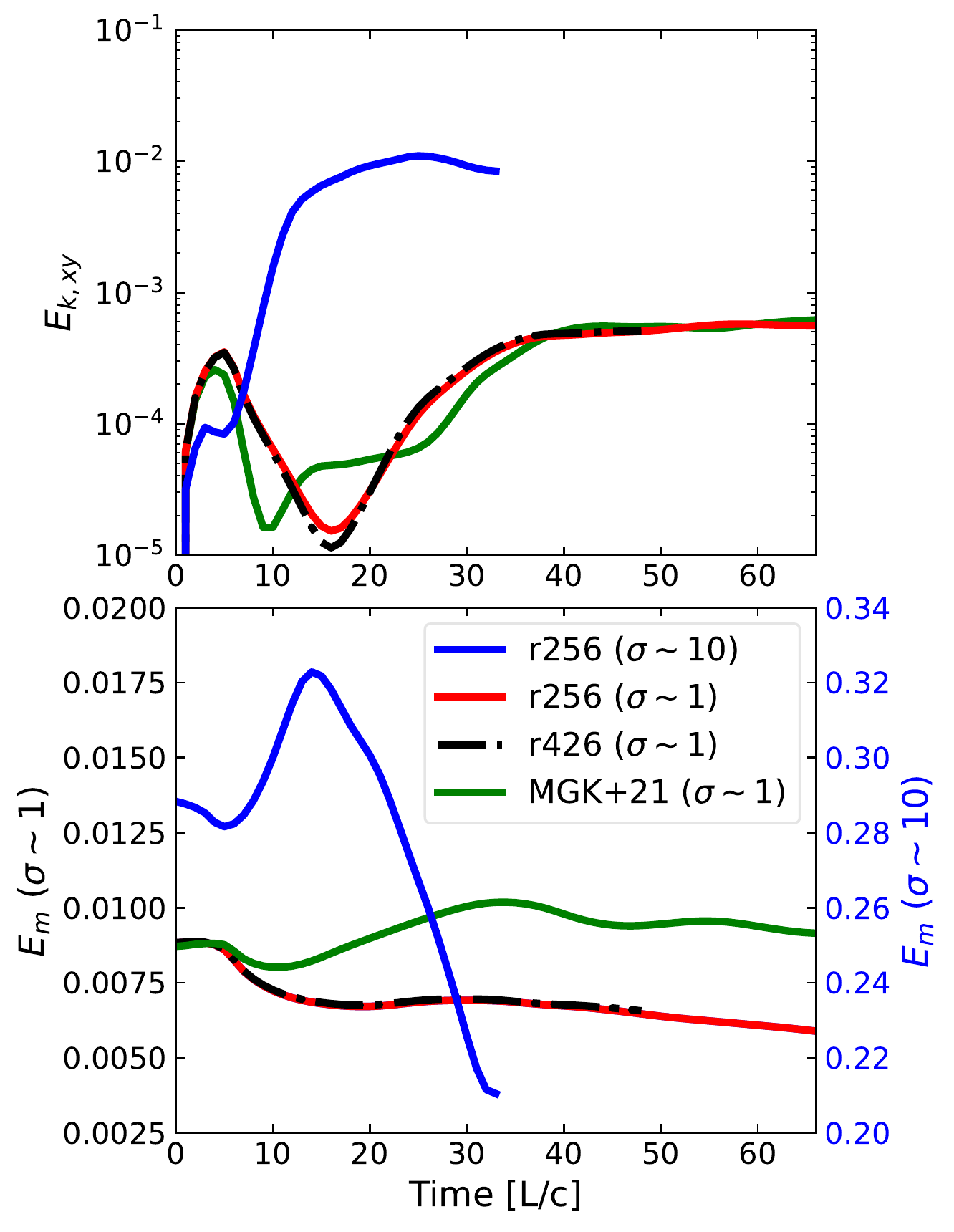}
\caption{Top: time evolution of the  volume-averaged kinetic energy density transverse to the z-axis within a cylinder of radius $R \leq 3.0L$  for the $\sigma \sim 1$ jet (red solid line for the model with resolution $256^3$  and dashed-dotted black line   for the model with resolution $426^2-256$), and for the  $\sigma \sim 10$ jet (blue solid line). Bottom: volume-averaged relativistic electromagnetic energy density for the same models. For comparison, also plotted with dashed red lines are the results obtained in MGK+21 for the $\sigma \sim 1$. The kinetic energy is presented in log scale, while $E_m$ is in linear scale.
}
\label{energy}
\end{figure}
\begin{figure*}[h] 
 \centering
  \includegraphics[scale=0.8]{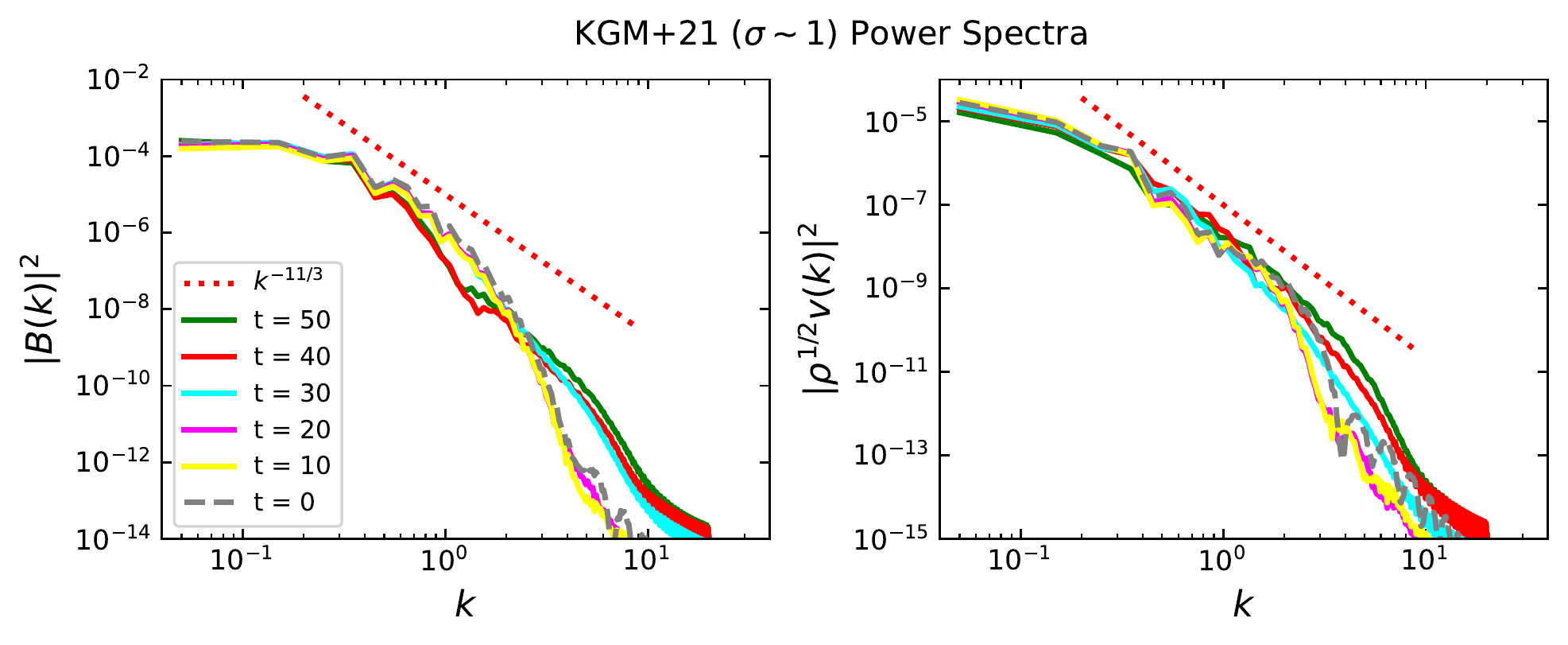}
  \includegraphics[scale=0.8]{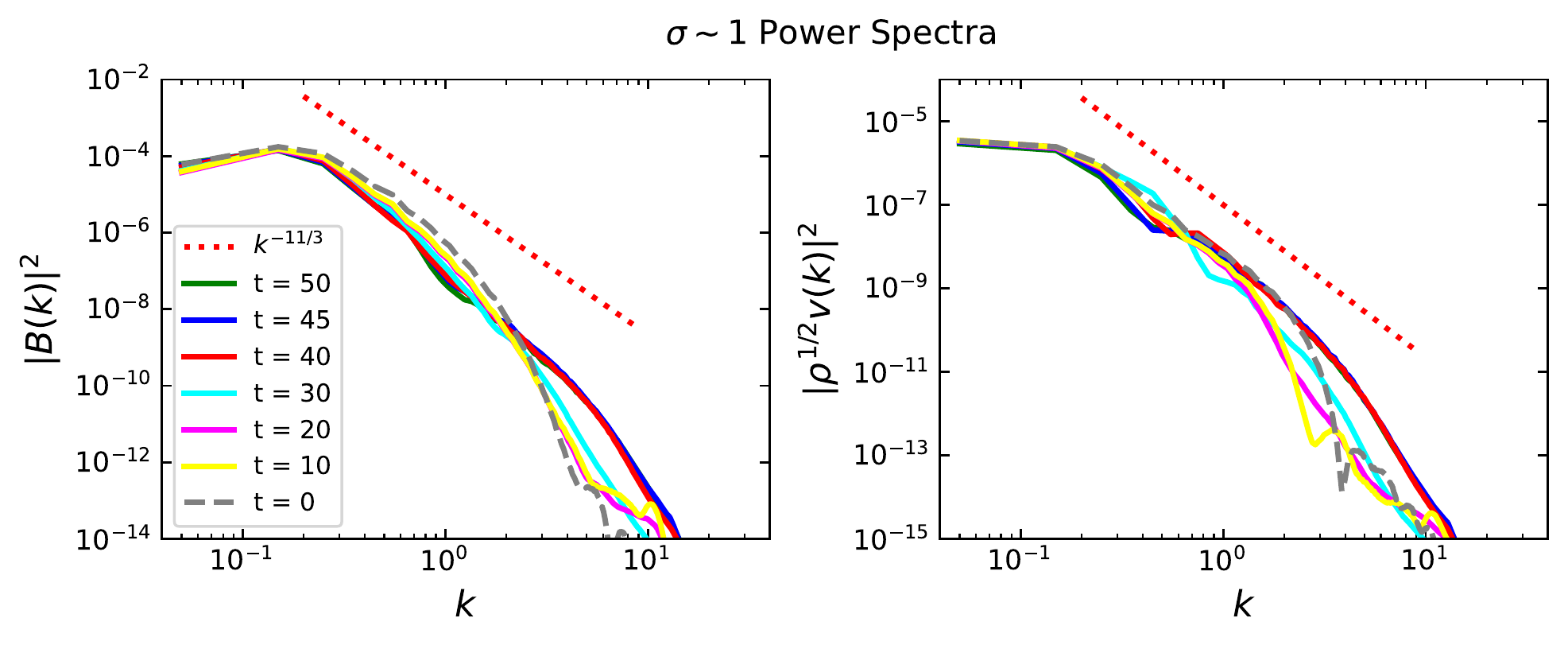}
  \includegraphics[scale=0.8]{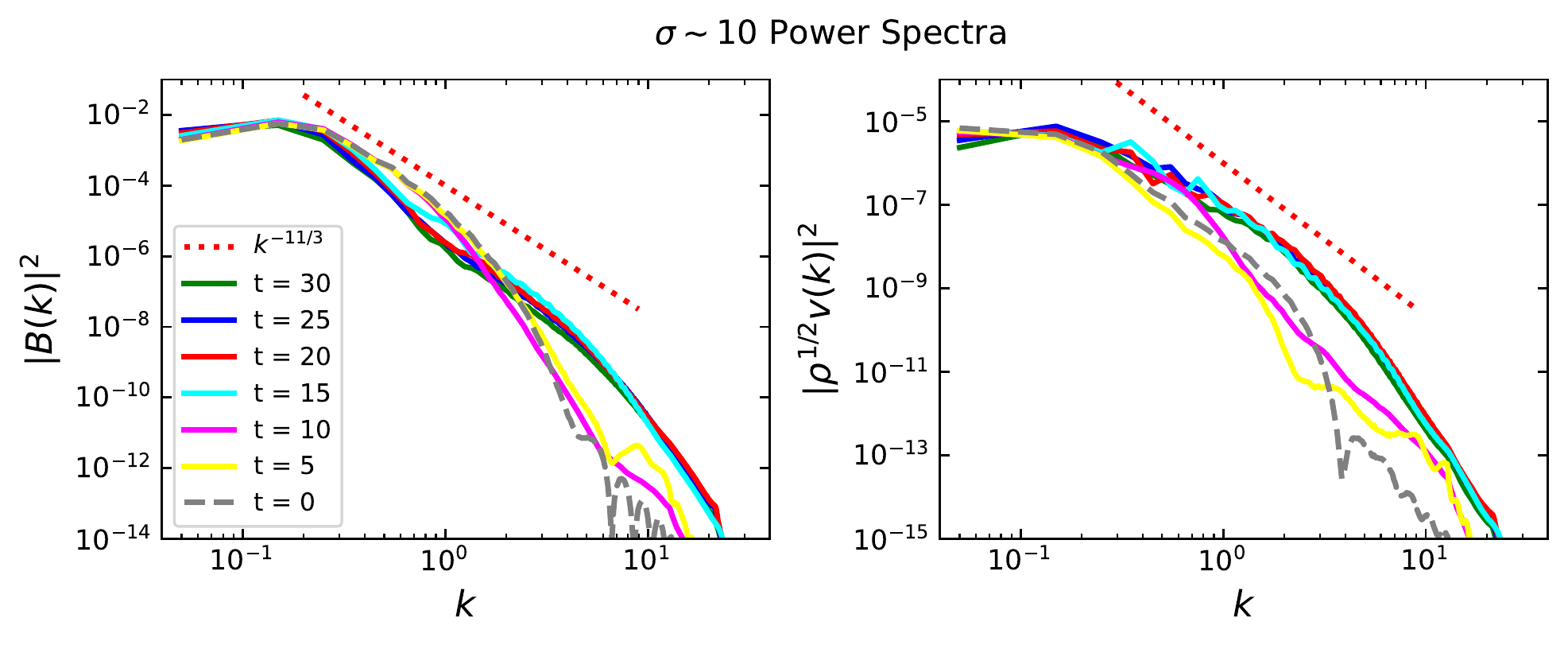}
\caption{Power spectrum  of the magnetic (left) and kinetic (right) energy densities  
for  the $\sigma \sim 1$ jet model of \citetalias{kadowaki_etal_2021} (upper row), the $\sigma \sim 1$  (middle row) and    $\sigma \sim 10$ (bottom row) jet models of this work,
for different times in unit of L/c.
The red doted line corresponds to a $k^{-11/3}$ 3D-Kolmogorov spectrum and its extension gives the inertial range of the turbulence for evolved times
$>30$ L/c
for the $\sigma \sim 1$ models, and $>10$ L/c for the $\sigma \sim 10$ model. The wavenumber is in unit of L.}
\label{spectrum}
\end{figure*}
\begin{figure} 
\centering
    \includegraphics[scale=0.4]{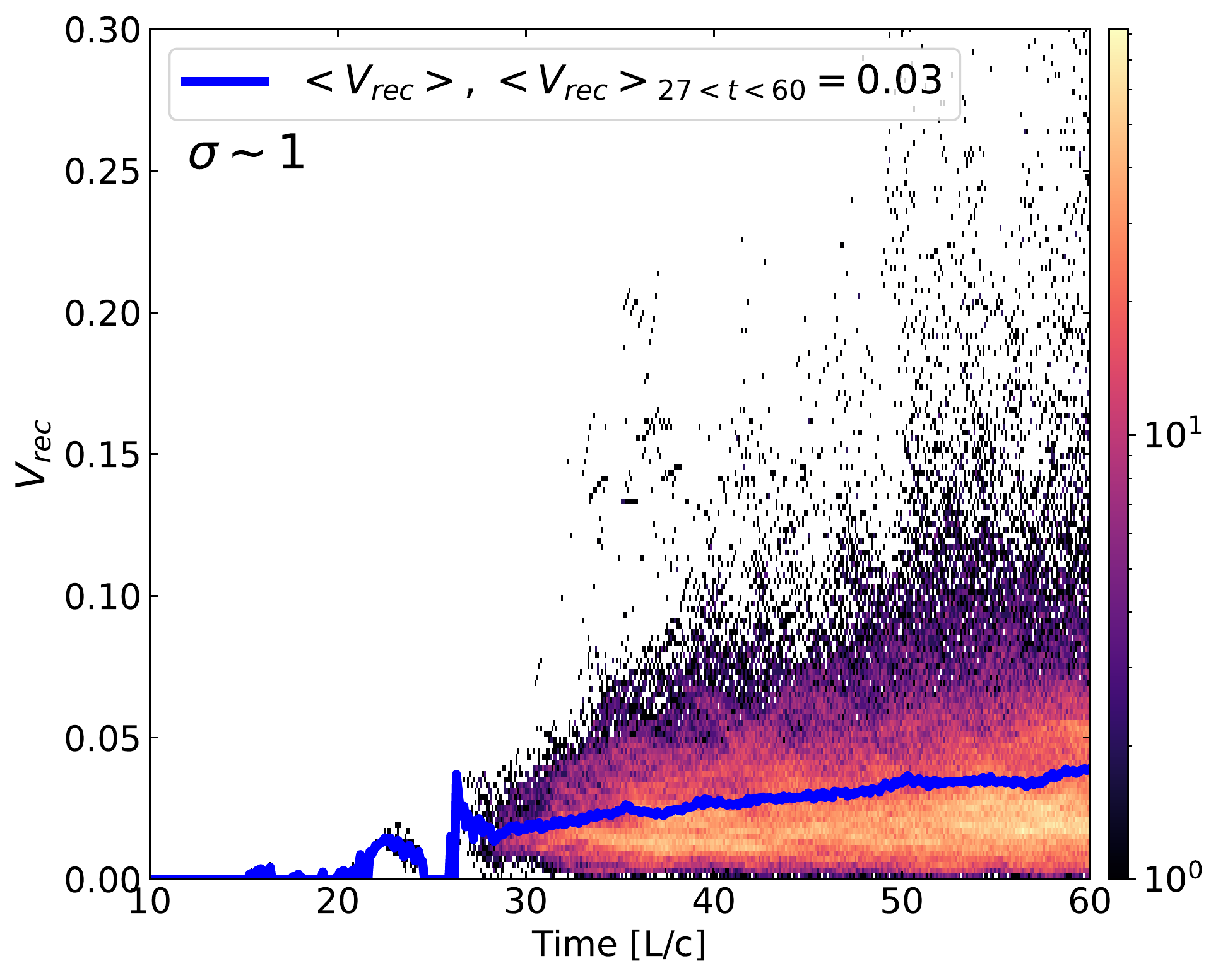} 
    \includegraphics[scale=0.4]{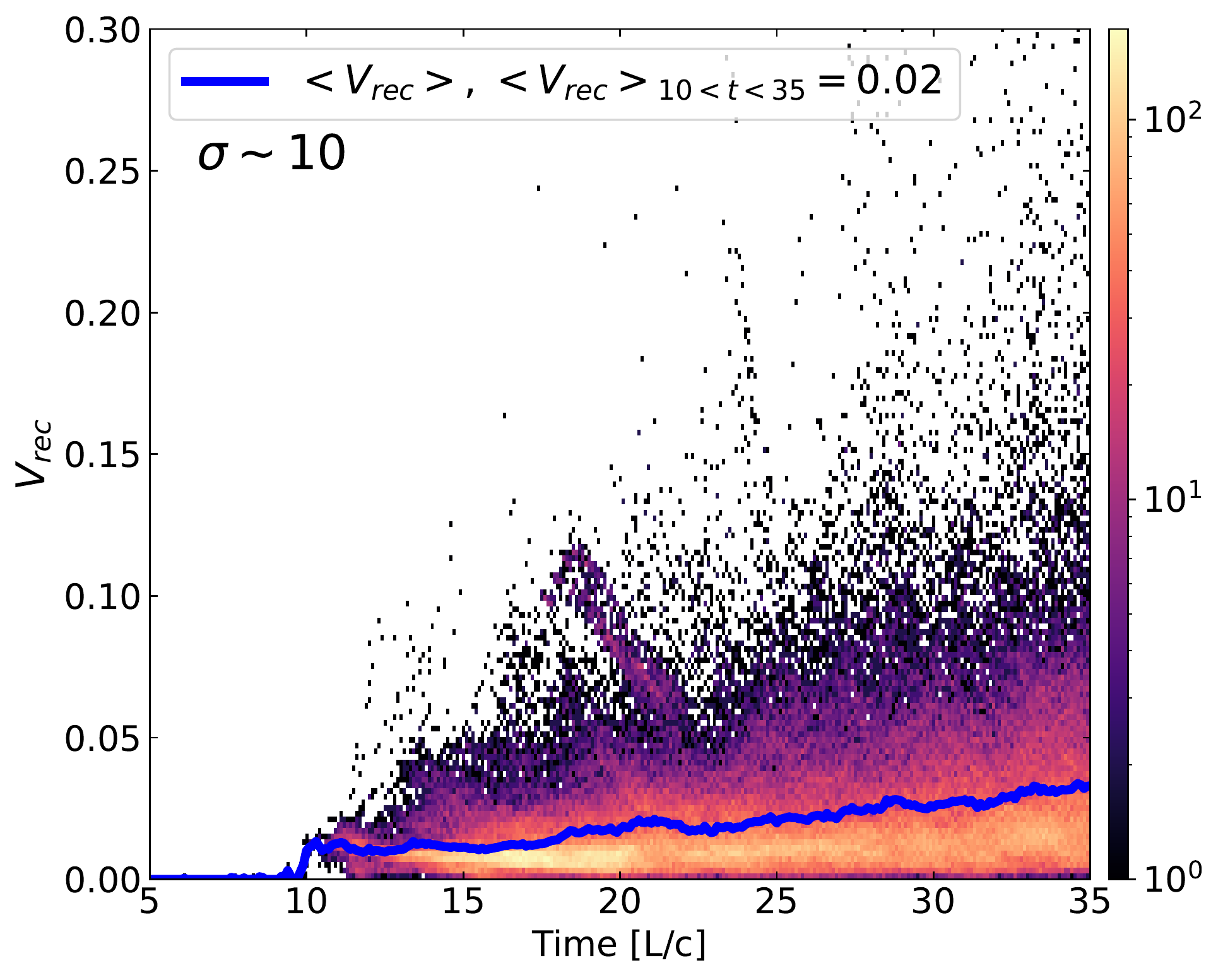} 
    \includegraphics[scale=0.4]{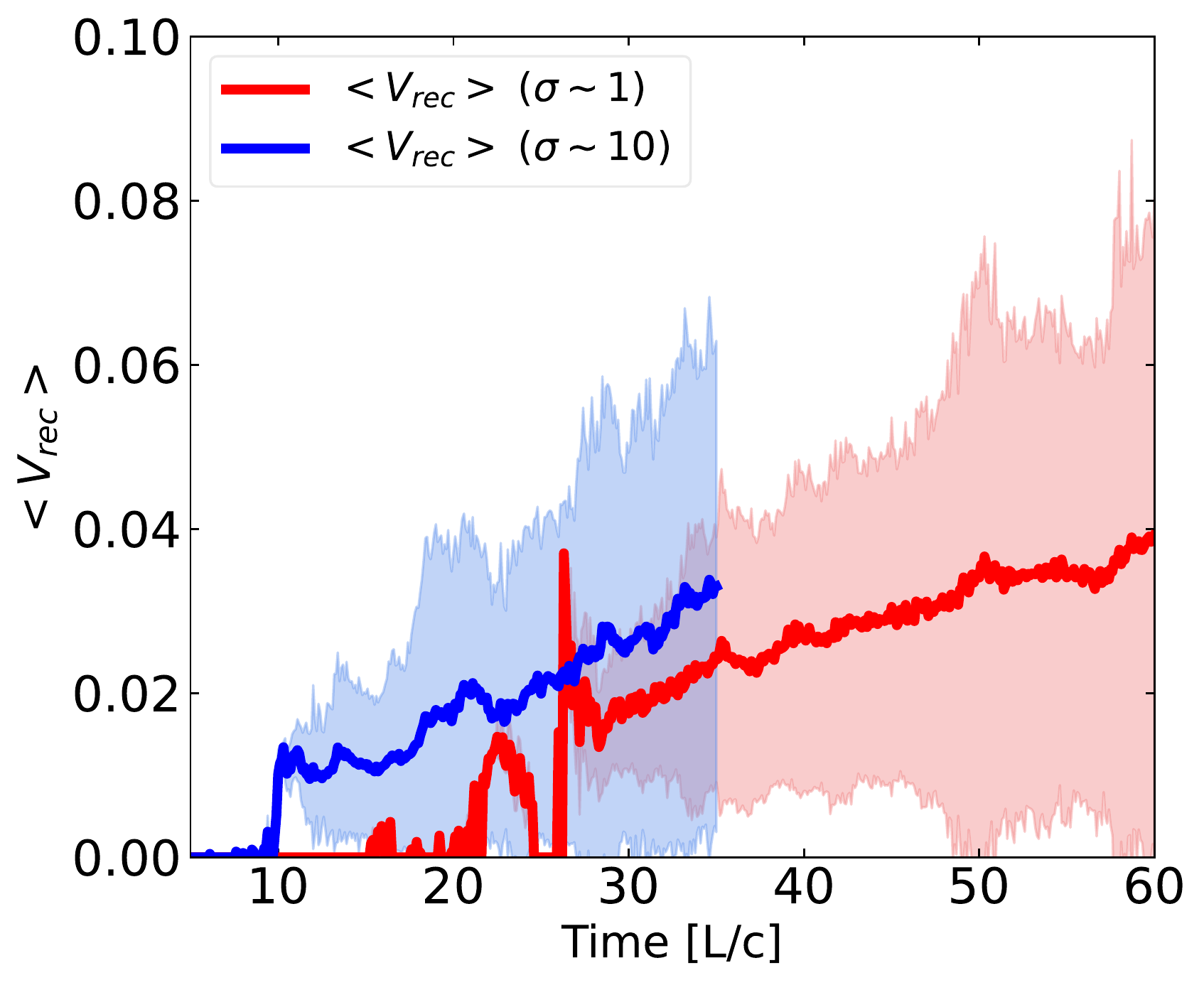}
\caption{Histogram of the reconnection rate evolution for the   $\sigma \sim 1$  (top) and  $\sigma \sim 10$ jet (middle). The blue line gives the average reconnection rate evolution. Bottom diagram compares the average reconnection rate evolution of the two models and the colored shades correspond to the standard deviations of each model.  
}\label{vrec}
\end{figure}

Figure \ref{energy} shows the time evolution of the volume-averaged kinetic energy density transverse to the z-axis (upper panel), 
and the volume-averaged total relativistic electromagnetic energy density ($E_m$) (bottom panel)  
for the $\sigma \sim 1$ jet, as the CDKI grows \citep[see also ][]{mizuno_etal_12,singh_etal_16,medinatorrejon_etal_2021}. 
For this jet model, these quantities are presented for two different resolutions, $256^3$ (solid red lines)  and $426^2 \times 256$ (dot-dashed black lines), and the results are both very similar.  
These curves are also compared with those  obtained by \citetalias{medinatorrejon_etal_2021} \citepalias[and][]{kadowaki_etal_2021} using the \texttt{RAISHIN} code for the same jet model
(labeled as MGK+21 in  Figure \ref{energy}),
and  with the $\sigma \sim 10$ jet. 
Note that $E_m$ is presented in the linear scale, while the kinetic energy is in the log scale.
The results of both $\sigma \sim 1$ jet models are comparable. 
As the CDKI develops,  $E_m$ is converted into kinetic energy.  For the $\sigma \sim 1$ models, the initial relaxation of the system to equilibrium leads to a hump in the kinetic and $E_m$ curves.
After this relaxation,   there is an initial   growth of  $E_m$  
caused by the increasing wiggling distortion of the magnetic field structure in the jet spine due to the initial growth of the CDKI. The kinetic energy, after a slower increase,  undergoes an exponential growth which is a little more advanced in time in the \texttt{PLUTO} run, that starts  around  $\sim  25$ L/c, than in the \texttt{RAISHIN} run (MGK+21), that starts around    $\sim  30$ L/c.  
This causes   the  jet model in this work to achieve earlier a turbulent state than in the model of MGK+21, with a time delay  $\Delta t \sim  5$ L/c between them\footnote{We attribute this small delay to intrinsic numerical differences between the two codes and to the slight difference in the grid resolution. The  $\sigma \sim 1$ jet model run with the \texttt{RAISHIN} code by \citetalias{medinatorrejon_etal_2021} has a cell size  $\sim 0.03$ L in the three directions.}. 
After the exponential growth, the kinetic energy reaches  approximately a plateau while $E_m$ decreases. 
This  coincides with full increase of the turbulence  and of the number of fast reconnection events in Figure \ref{jet_points} (bottom right; see also Figure \ref{vrec}). In fact, this plateau characterizes the achievement of  saturation  of the CDKI and a nearly steady-state turbulent regime in the system (see Figure \ref{spectrum}). A similar behaviour has been identified in \citetalias{medinatorrejon_etal_2021} and \citetalias{kadowaki_etal_2021}. We also notice that there is a  difference of at most  $30\%$ in the amplitude  of $E_m$ between the two models.
 In the  $\sigma \sim 10$ jet, the CDKI clearly increases faster  achieving saturation much earlier, at about half of the time of the $\sigma \sim 1$ jet. 
 
 Since the two models with different resolution for the $\sigma \sim 1$ jet are so similar,  in the rest of the manuscript we consider only the $256^3$ resolution model.
 




To  quantify the development  of the turbulence, 
we have evaluated the three-dimensional power spectra of the magnetic and kinetic energy densities in the  jet, considering averages in  
spherical or ellisoidal 
 shells between $k$ and $k + dk$ (where $k=\sqrt{k_x^2+k_y^2+k_z^2}$ in the Fourier space) \citepalias{kadowaki_etal_2021}. Figure \ref{spectrum} depicts 
 these power  spectra
 for different 
 times 
 for both,   $\sigma \sim 1$ and $\sigma \sim 10$ jets. A $3D$-Kolmogorov spectrum slope  ($\propto k^{-11/3}$; red dotted line) was included for comparison. 
 The diagrams show inertial ranges 
 both for the kinetic    $|\sqrt{\rho}\boldsymbol{v}(\boldsymbol{k})|^2$ 
 and 
 for the magnetic  $|\boldsymbol{B}(\boldsymbol{k})|^2$ energy density spectra
 between $0.2 \lesssim  k \lesssim 25$ (in units of 1/L) 
 in agreement with a Kolmogorov-like spectrum, after $t\simeq30$L/c for the $\sigma \sim 1$ jet and $t\simeq10$L/c for the $\sigma \sim 10$ jet. This indicates a turbulent energy cascade between an injection scale $\sim 5$L and a resistive small scale $\sim 0.11$L.  The magnetic energy spectrum shows a little steeper slope, probably due to the strong (guiding) magnetic  field of the background plasma \citep[see, e.g.,][]{kowal_etal_07,kadowaki_etal_2021}. As expected, the $\sigma \sim 10$ jet has maximum magnetic energy density 10 times larger than the $\sigma \sim 1$ jet.
 The results are comparable to those obtained in 
\citetalias{kadowaki_etal_2021} for the $\sigma \sim 1$ jet, as shown in the left diagrams of the figure\footnote{We note that the  turbulent power spectra  of the kinetic and magnetic energy densities   of the $\sigma \sim 1$ jet presented in \citetalias{kadowaki_etal_2021} were produced with  a distinct  normalization from the one used in Figure \ref{spectrum}. For this reason, we have reproduced them again here for direct comparison with  the other spectra of  Figure \ref{spectrum}.}. 

In order to identify fast magnetic reconnection sites in the turbulent flow  of the relativistic jet  and quantify their reconnection velocities, we have used the same  algorithm employed in \citetalias{kadowaki_etal_2021} wherein the method is described in detail  \citep[see also,][]{zhdankin_etal_13,kadowaki_etal_2018b}.
The time evolution of the  magnetic reconnection rate, ${V}_{rec}$, for all identified sites  and  the  time evolution of the average value, $\langle {V}_{rec}\rangle$ (blue line in the upper and middle panels), in units of the Alfvén velocity, are shown in Figure \ref{vrec}.
The evolution of $\langle{V}_{rec} \rangle$   changes more abruptly after  $t \sim25$ in the $\sigma \sim 1$  jet and $t \sim 10$ in the $\sigma \sim 10$  jet, when the CDKI starts to grow exponentially (Figure \ref{energy}). After that, as the CDKI tends to saturation, the average reconnection rate also attains a value $\langle{V}_{rec} \rangle \sim 0.03 \pm 0.02$ for the $\sigma \sim 1$ jet, in agreement with \citetalias{kadowaki_etal_2021} (see their reference model  m240ep0.5 and their Figure 8). For the $\sigma \sim 10$ jet, it is still growing to a plateau to a similar average value (middle diagram) $\langle{V}_{rec} \rangle \sim 0.02 \pm 0.02$.
A peak reconnection rate of the order $\sim 0.9$ (not shown in the figure)
is obtained for the  $\sigma \sim 1$ jet, 
while a peak value $\sim 0.6$ is attained for the $\sigma \sim 10$ jet. The bottom diagram compares directly the evolution of the average reconnection speed of both models including their respective variances which are similar\footnote{We note that the slightly smaller mean value of the reconnection rate for the larger $\sigma$  model is compatible with the fact that the necessary wandering of the field lines by the turbulence in order to drive fast reconnection is naturally more difficult the larger the strength of the magnetic field \citep{lazarian_vishiniac_99}.}.
 


\begin{figure} %
  \centering
   \begin{overpic}[scale=0.44]{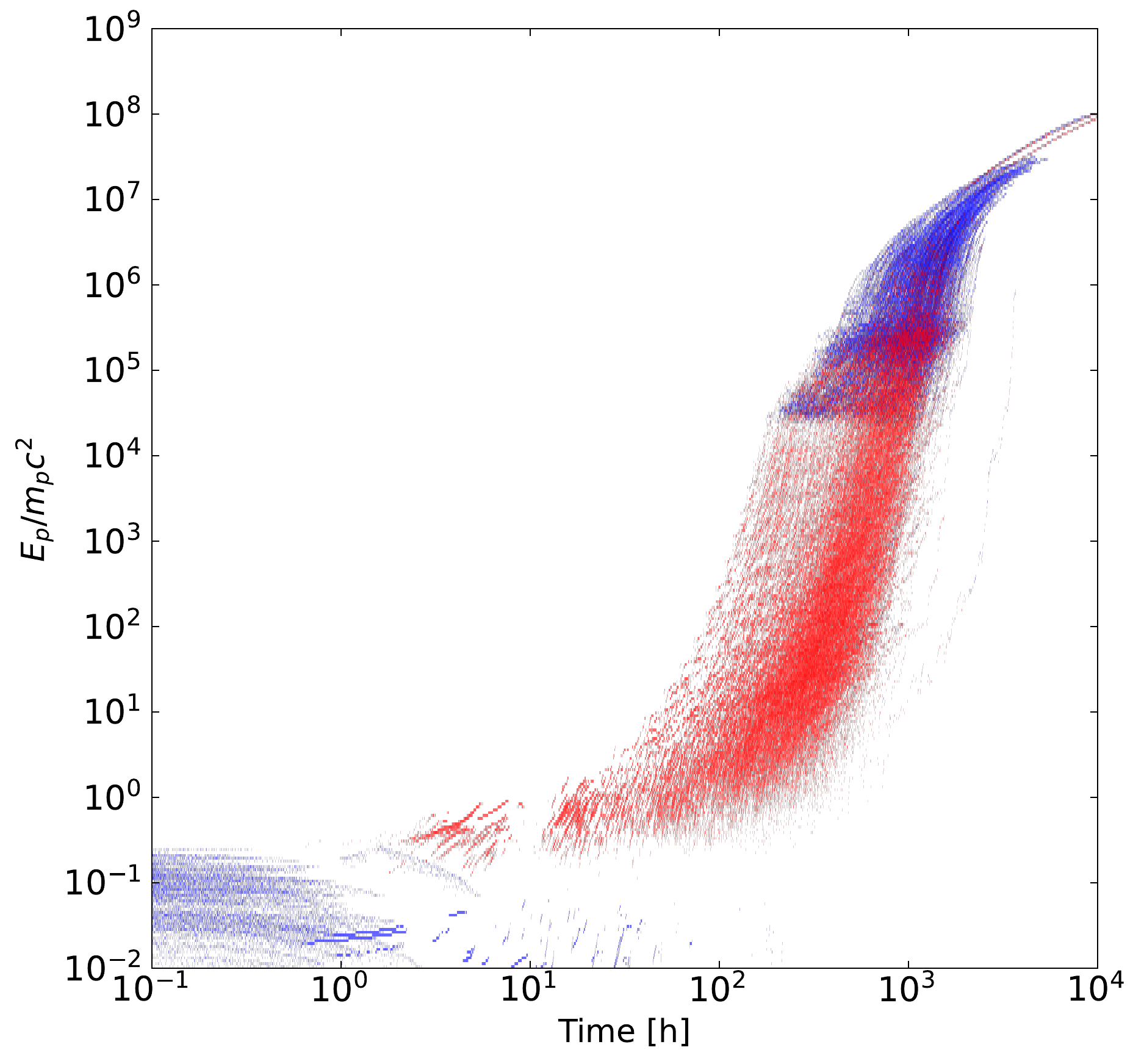}
    \put(14,50){\includegraphics[scale=0.23]{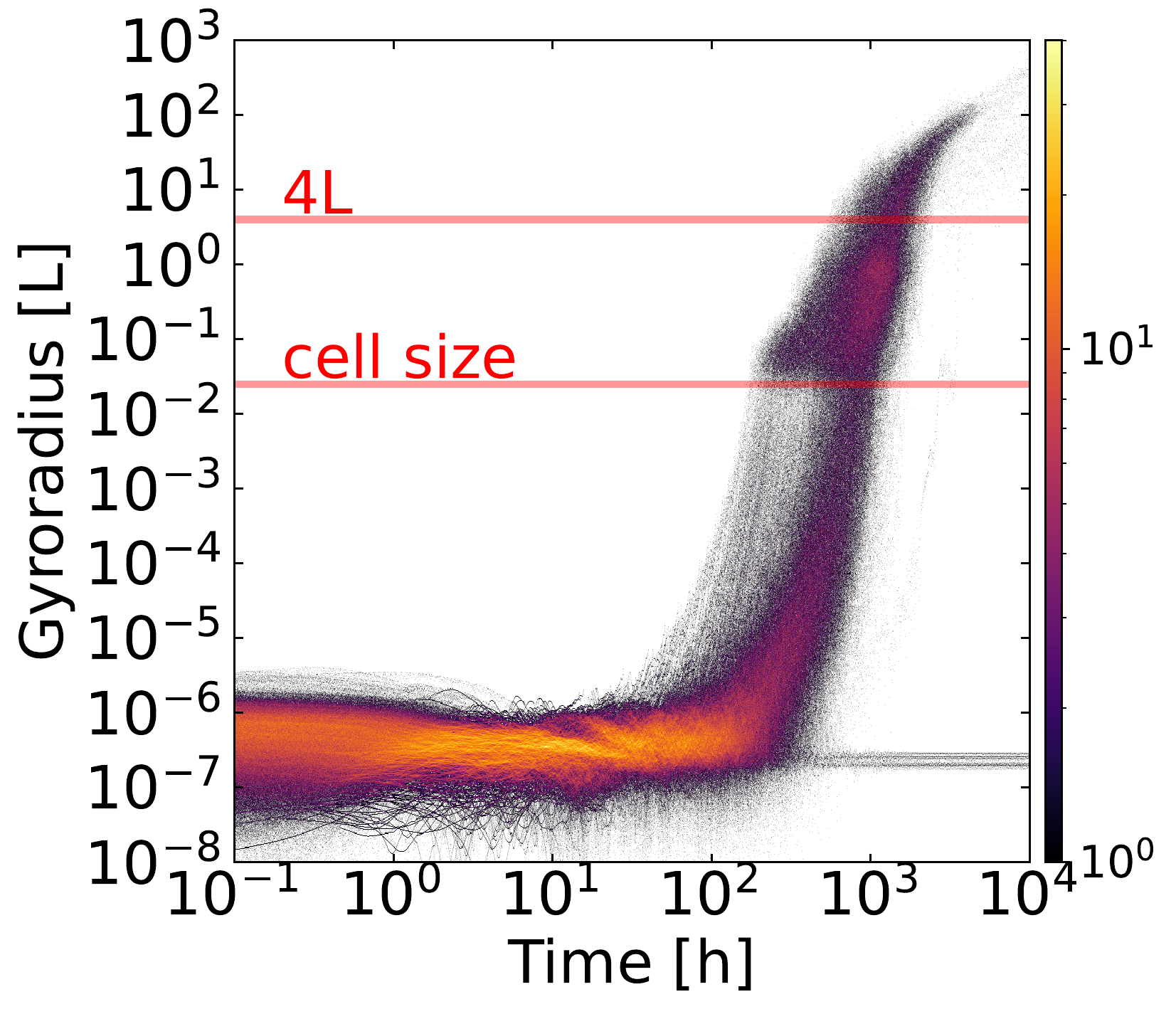}}
  \end{overpic} 
  \begin{overpic}[scale=0.44]{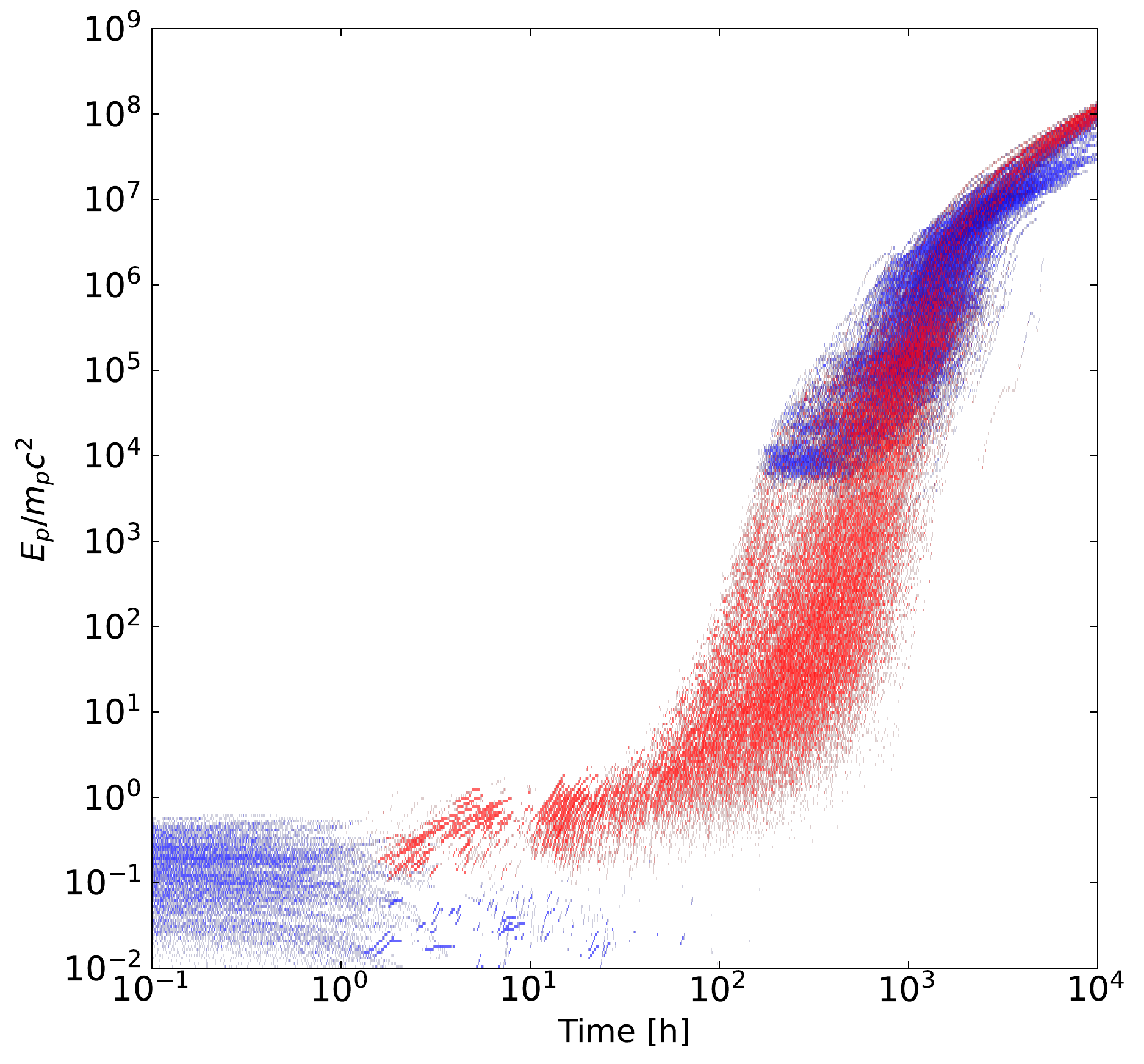}
    \put(14,50){\includegraphics[scale=0.23]{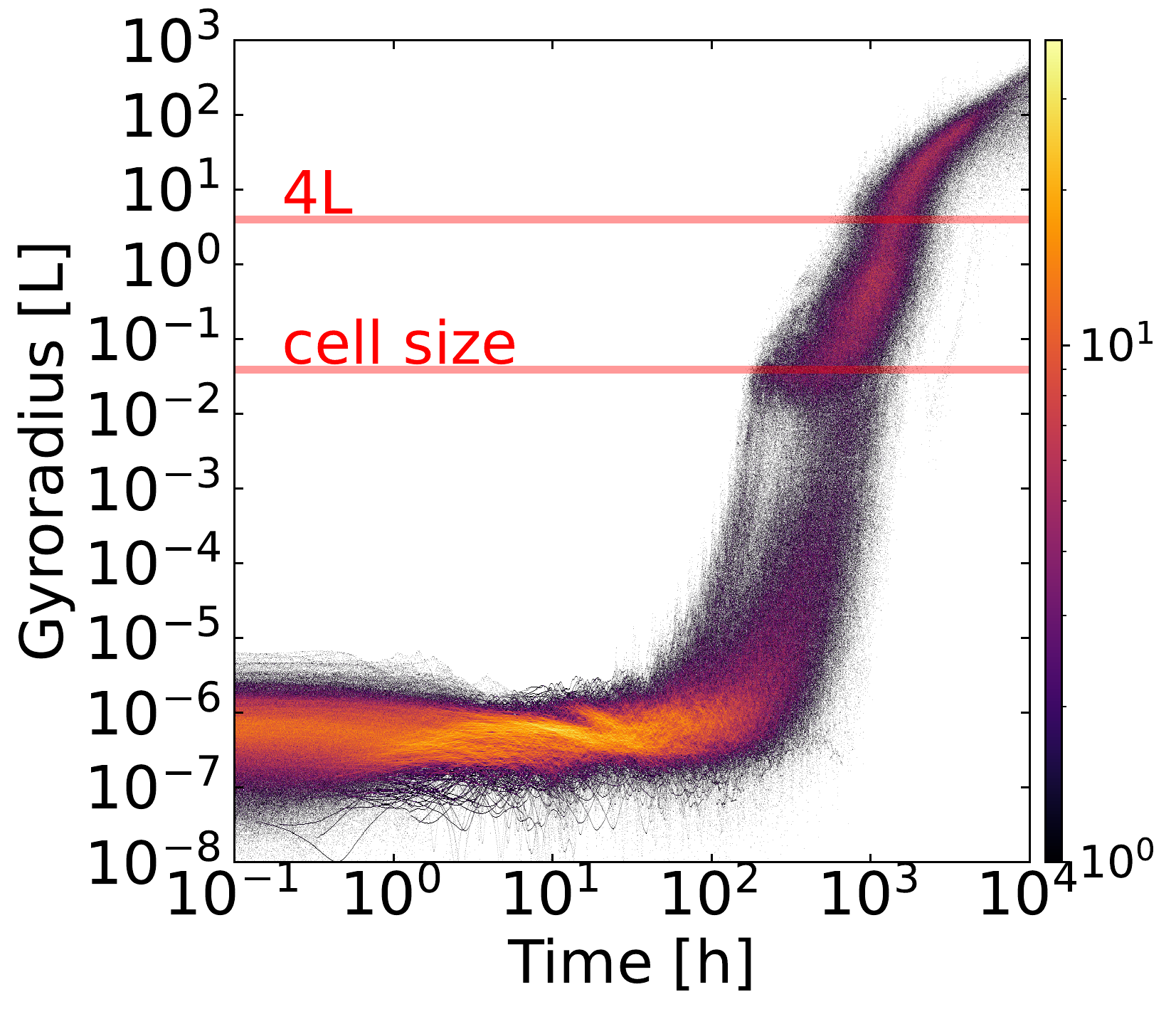}}
  \end{overpic}  
\caption{Kinetic energy evolution, normalized by the proton rest mass energy, for 1,000  particles injected into the fully turbulent snapshot $t =50$ L/c  of  the $\sigma \sim 1$ jet run by \citetalias{medinatorrejon_etal_2021} (top). The same for particles injected into the the snapshot $t =45$ L/c of the $\sigma \sim 1$ jet in this work (bottom). 
 The colors indicate which velocity component is being accelerated (red or blue for the parallel or perpendicular component to the local magnetic field, respectively). The insets in the upper left corner show the time evolution of the particles gyroradius. The color bars indicate the number of particles. The horizontal grey stripe  is bounded on the upper part by the jet diameter  ($4L$) and on lower part by the cell size of the simulated background jet. In these  particle simulations, the particle acceleration  time  is given in hours and the adopted physical size for $L$  is the same as in \citetalias{medinatorrejon_etal_2021} for comparison, $L = 3.5 \times 10^{- 5}$ pc.
}
\label{GACELL-PLUTO45}
\end{figure}

In \citetalias{medinatorrejon_etal_2021}, test particles were injected with an initial Mawellian distribution (with initial mean kinetic energy $\left < E_p \right > \sim 10^{-2} m_p c^2$) in the simulated  $\sigma \sim 1$  jet  with already fully developed turbulence (with the \texttt{RAISHIN} code), and accelerated by magnetic reconnection up to VHEs. 
Figure \ref{GACELL-PLUTO45} (upper panel) depicts the kinetic energy growth as a function of time for 1,000 particles injected (with the \texttt{GACCEL} code) in the snapshot $t=50$ L/c of their model \citepalias[see also  bottom panel of Figure 5 in][]{medinatorrejon_etal_2021}. 
The lower panel of Figure \ref{GACELL-PLUTO45}  shows  a similar  plot, but obtained  for particles injected (also with the \texttt{GACCEL} code) in the fully turbulent jet simulated in this work with the \texttt{PLUTO} code, at $t=45$ L/c. As remarked previously in Figure \ref{energy}, the model run here develops turbulence earlier, with an advance in time of $\Delta t \sim 5$ L/c and thus, in order to compare with \citetalias{medinatorrejon_etal_2021} results, we have considered the corresponding earlier snapshot.   The results are very similar, as expected. As in \citetalias{medinatorrejon_etal_2021},  particles are accelerated exponentially in the  magnetic reconnection sites in all scales of the turbulence driven by the CDKI up to $\sim 10^7 m c^2$, which corresponds to a Larmor radius comparable to the  diameter of the  jet and the size of the largest turbulent magnetic structures (see the plot in the inset). As we see in the figure, beyond this energy, particles suffer further acceleration at a smaller rate, which is attributed to drift in the large scale non-reconnected fields.  We also see that the parallel component of the velocity is predominantly accelerated in the exponential regime, as expected in a Fermi-type process, while in the drift regime, it is the perpendicular component that prevails (see \citetalias{medinatorrejon_etal_2021} for more details).


The figures described above  evidence the similarity of the results obtained with the two MHD codes and reinforce the results of \citetalias{medinatorrejon_etal_2021} and \citetalias{kadowaki_etal_2021}.

\begin{figure} %
  \centering
  \begin{overpic}[scale=0.44]{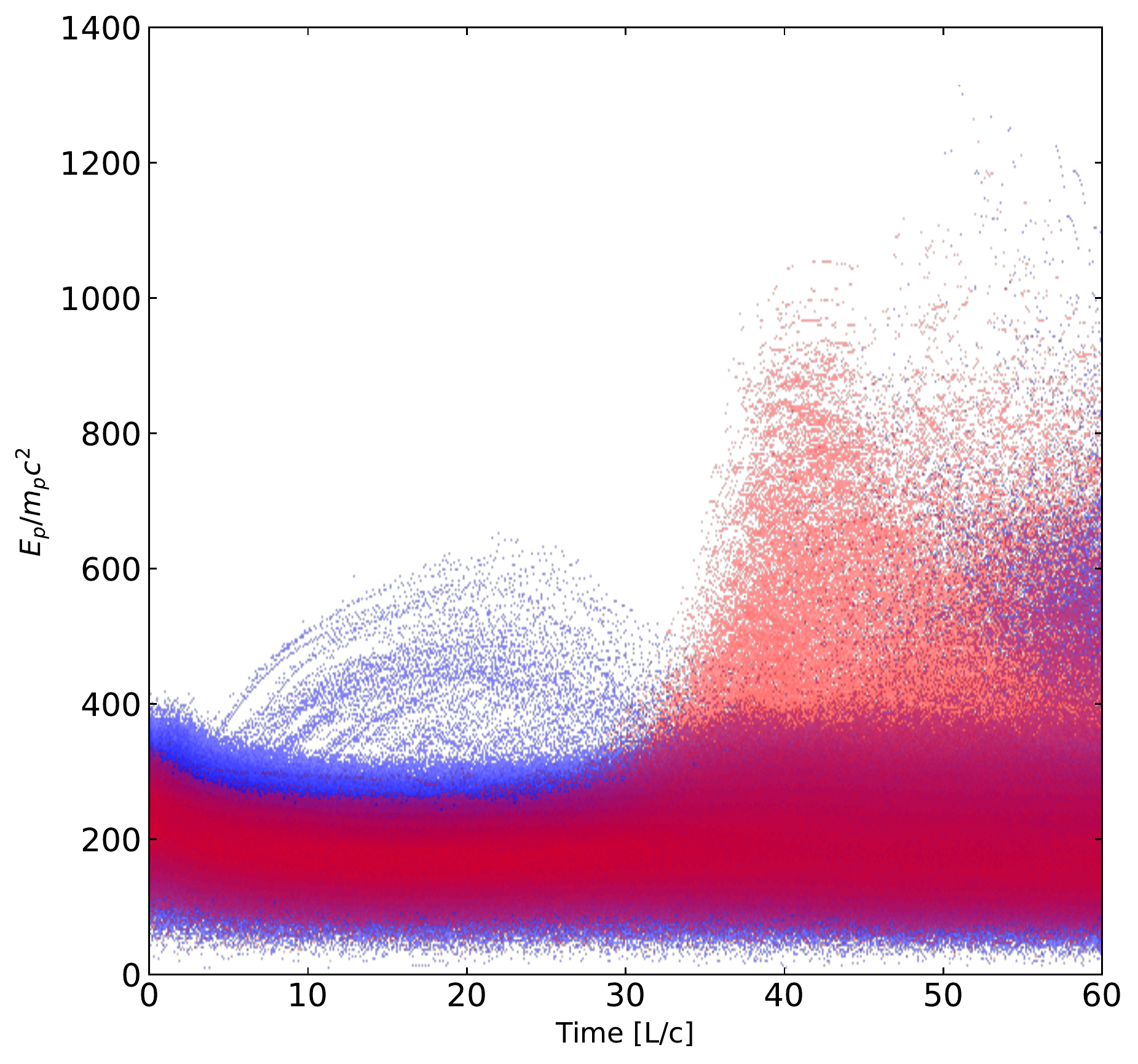}
    \put(14,52 ){\includegraphics[scale=0.19]{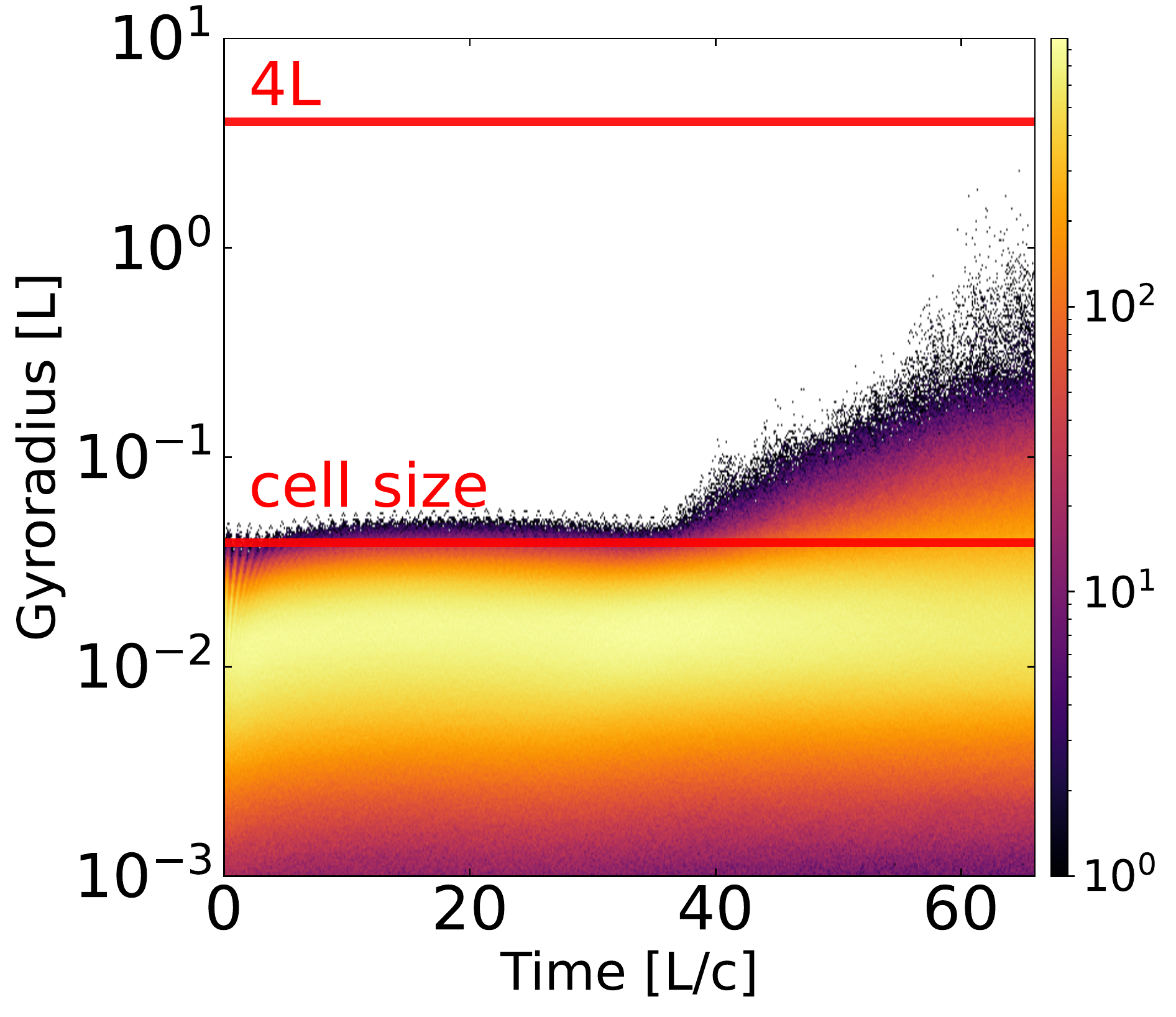}}
  \end{overpic} 
  \begin{overpic}[scale=0.44]{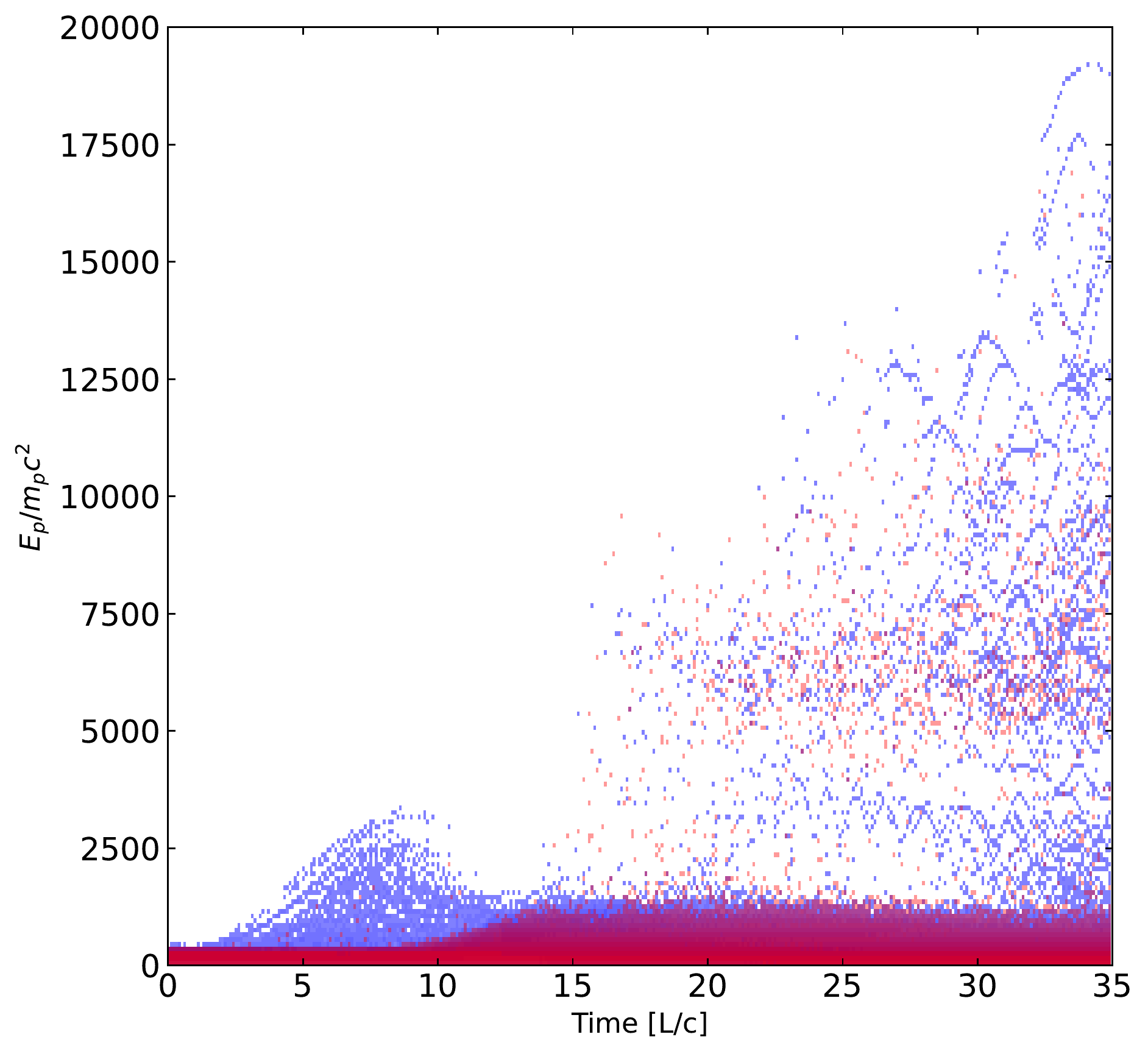}
    \put(15,51){\includegraphics[scale=0.19]{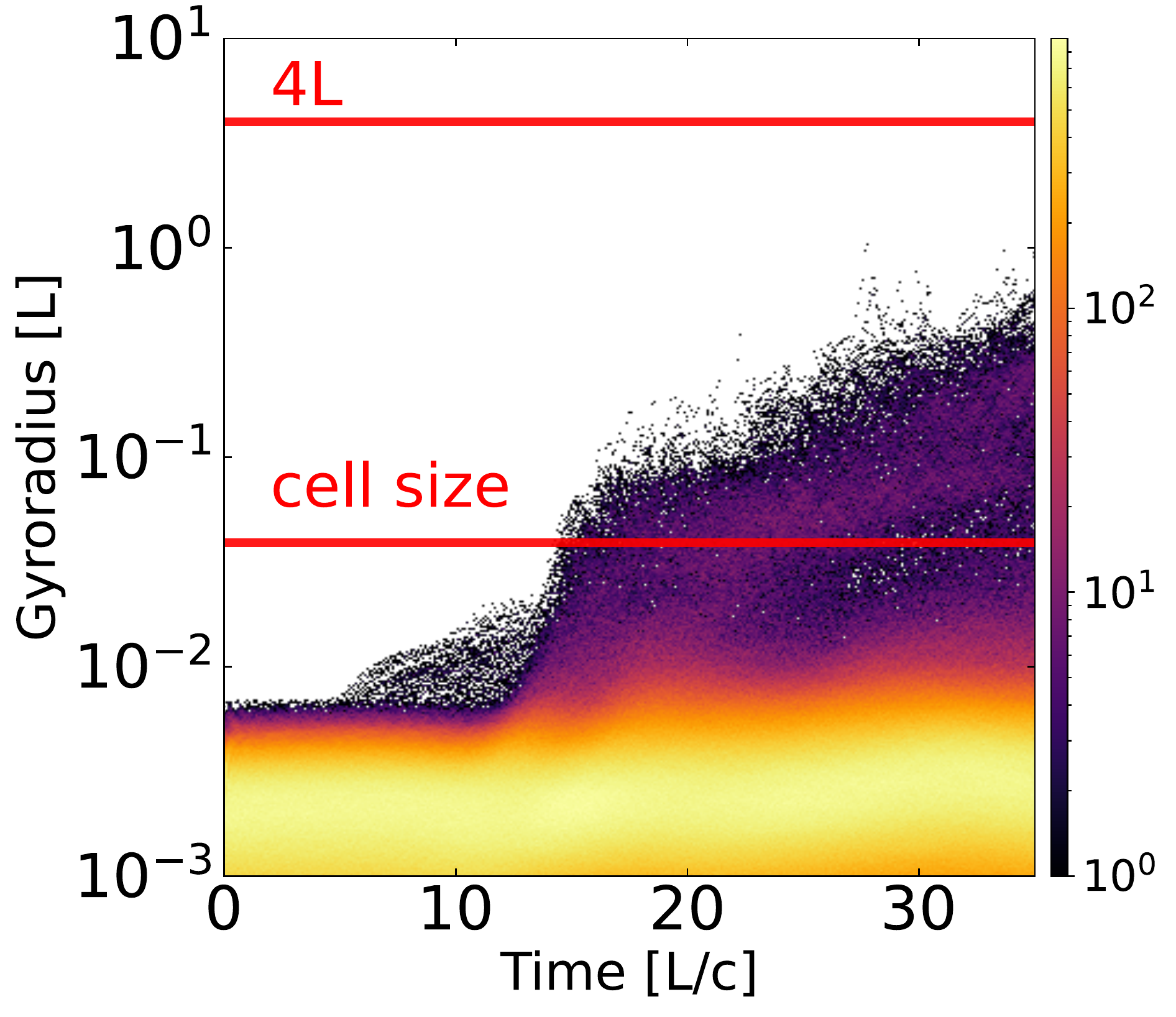}}
  \end{overpic}  
\caption{Kinetic energy evolution for 50,000 particles evolved in the MHD-PIC simulation for the $\sigma \sim 1$ (top) and for the  $\sigma \sim 10$ (bottom) jet. Particles are initially injected with  energy $\left < E_p \right > \sim 1-200 m_p c^2$. 
The colors indicate which velocity component of the particles is being accelerated (red or blue for the parallel or perpendicular component to the local magnetic field, respectively). The inset panels depict the evolution of the particles gyroradius, and the red horizontal lines correspond to the jet diameter ($4L$) (top)  and the cell size of the simulated jet (bottom).
}
\label{pic1}
\end{figure}

Figure \ref{pic1} shows the first stages of the kinetic energy evolution of the particles evolving together with the background  jet as obtained with the present 
model (i.e., employing the MHD-PIC mode) both for the $\sigma \sim 1$ and $\sigma \sim 10$ jet.  In the very beginning, while the CDKI is still developing, particles only suffer drift in the background magnetic fields. Then, as the jet column  starts to wiggle  around $t \sim 20$ L/c in the $\sigma \sim 1$, and around $t \sim 7$ L/c in the  $\sigma \sim 10$ jet, due to the kink instability (Figure \ref{jet_points}),  the particles suffer curvature drift acceleration. Note that at these times, fast reconnection driven by turbulence is not developed yet (Figure \ref{vrec}). As stressed earlier, curvature drift  acceleration has been also detected in the $\sigma \sim 1$ jet by \citetalias{medinatorrejon_etal_2021}, for a similar resolution,  around similar jet dynamical time (more precisely, at $t\sim 25$  L/c, due to the time delay between the two runs; see their Figure 6), and by \citet{alves_etal_2018} in PIC simulations of the early stages  of the development of the kink instability. 

After $t \sim 30$ L/c in the $\sigma \sim 1$ jet  (and $t \sim 15$ L/c in the $\sigma \sim 10$ jet), which coincides with the nonlinear growth  and saturation  of the CDKI leading to fully developed turbulence in the jet (Figures \ref{energy} and \ref{spectrum}), the particles in Figure \ref{pic1} start exponential acceleration, 
as in Figure \ref{GACELL-PLUTO45}. The maximum achieved energy is about 10 times larger for the jet with  corresponding larger  $\sigma.$ 
The entire dynamical time of the system evolution is of only $60$ L/c for the   $\sigma \sim 1$ jet (and half this time for the $\sigma \sim 10$ jet). For the particles, the physical time elapsed is  only  $\sim 60 L/c \sim 1$ hr (and half-hour for the  $\sigma \sim 10$ jet, for the adopted $L=5.2 \times 10^{- 7}$  
pc in physical units), which is much smaller than the several hundred hours that particles can accelerate in the nearly steady state jet snapshot of Figure \ref{GACELL-PLUTO45}  where they can re-enter the system several times through the periodic boundaries of the jet in the z direction until they reach the saturation energy (see also \citetalias{medinatorrejon_etal_2021}).
This explains why particles do not achieve the maximum possible energy
by acceleration in the largest turbulent magnetic reconnection structures of the order of the jet diameter ($\sim  4L$), as we see in the inset in the figure, which depicts the particles Larmor radius distribution. 
For this value of the Larmor radius ($R_{max}\sim 4L$), the particles would achieve an energy $E_{sat} \sim e \, B \, R_{max} \sim 200,000$ $m_p c^2$ in the  $\sigma \sim 1$ jet, and $\sim 1,000,000$ $m_p c^2$  in the $\sigma \sim 10$ jet, if  the jet were allowed to evolve for a dynamical time about one hundred times larger (where $R_{max} \sim 4L = 2.1 \times 10^{-6}$ pc, and $B \sim 0.1$ G and $\sim 0.6 $ G for the $\sigma \sim 1$ and $\sigma \sim 10$ jets, respectively, considering the physical units employed in the MHD-PIC simulations).  
Nonetheless, the results in these early stages of particle acceleration,  follow the same trend depicted in Figure \ref{GACELL-PLUTO45}, indicating that particles are accelerated exponentially by magnetic reconnection in the turbulent flow, from the small resistive scales up to the large scales of the turbulence in the ideal electric field of the magnetic reconnecting structures. These results also indicate that the time evolution of the background magnetic fields does not influence the acceleration of the particles since they enter the exponential regime of  acceleration in the same jet dynamical times in which turbulence becomes fully developed, as obtained  in the MHD simulations with test particles of Figure \ref{GACELL-PLUTO45}. At the more evolved dynamical times, particularly in the $\sigma \sim 10$  jet, we also identify particles having their perpendicular velocity component being accelerated suggesting the presence of drift acceleration too, as in the late stages of particle acceleration in  Figure \ref{GACELL-PLUTO45}.

We have also run the MHD-PIC model for the $\sigma \sim 1$ and 10 jets with the larger resolution $426^2-256$,
and the results we obtained for particle acceleration evolution are very similar to those shown in Figure \ref{pic1}. 
The only difference is that less particles re-enter the system and thus the histogram has comparatively less accelerated particles. In particular, there are  almost no particles undergoing curvature drift in the very early times  (around $t\sim 20$ L/c), 
but the exponential  regime, with a dominance of  the acceleration of the parallel component of the velocity, is clearly detected, as in Figure \ref{pic1} (top)\footnote{The absence of accelerated particles by curvature drift in this case could be explained by the fact that this acceleration can be experienced only by particles with a Larmor radius large enough to $feel$ the curvature of the field (\citet{alves_etal_2018}, \citetalias{medinatorrejon_etal_2021}). When we increase the resolution  of the MHD domain (and thus decrease the cell size), particles with the same (still small) Larmor radius, at the same dynamical time step around $t\sim 20$ L/c  as in the lower resolution simulation (Figure \ref{pic1}), will see no field curvature when moving  from a smaller cell to the other and then, experience only linear drift, as in much earlier times.}.

\begin{figure} 
\centering
    \includegraphics[scale=0.45]{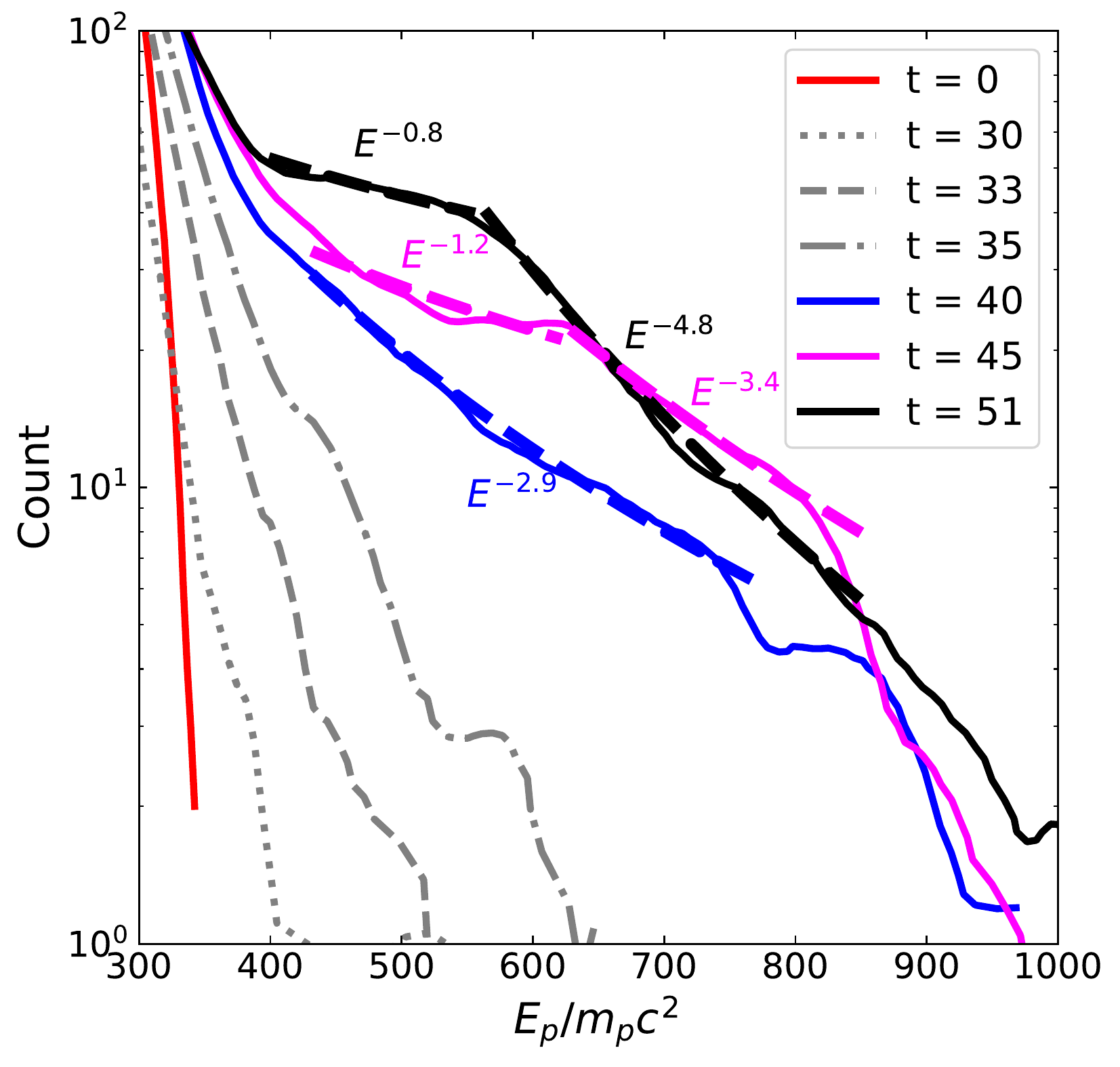}
    \includegraphics[scale=0.45]{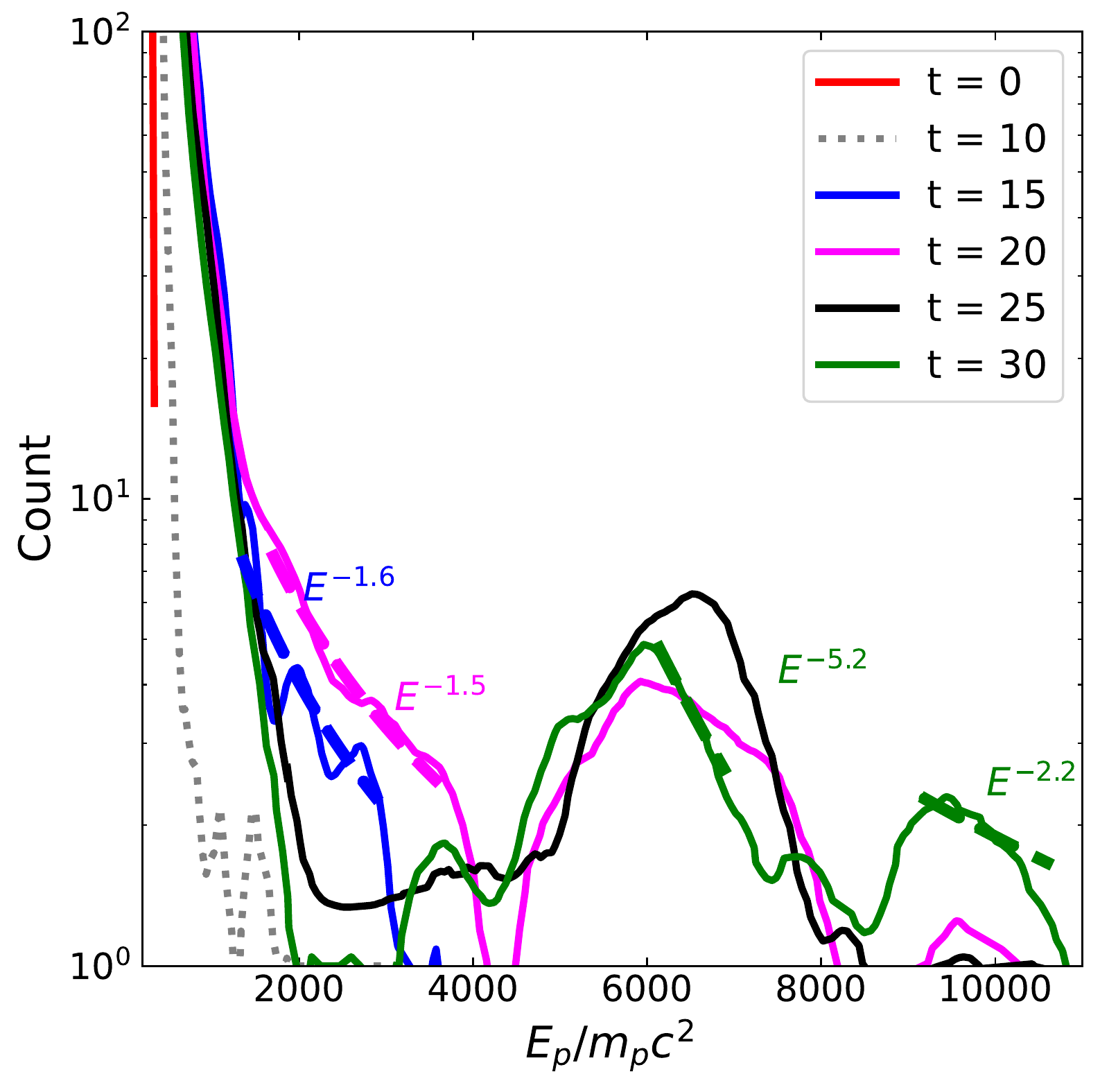}
\caption{Particle energy spectrum evolution as a function of the normalized kinetic energy  for the particles evolved in the MHD-PIC simulation for the $\sigma \sim 1$ (top)  and $\sigma \sim 10$  (bottom) jet. 
 The solid red line corresponds to the initial 
 distribution. The high-energy tails in more evolved times of the system are fitted by  power laws.
}
\label{picmhd-NE}
\end{figure}

In Figure \ref{picmhd-NE}  we show the  particle energy spectrum for the $\sigma \sim 1$ and $\sigma \sim 10$ jets, for different time steps in these early stages of the acceleration. 
The initial 
distribution 
is represented by a red line.
As particles accelerate, they start to populate  the high energy tail in the distribution, which becomes flatter  as time evolves. In the $\sigma \sim 1$ jet, 
we note the formation of two slopes in more evolved times with a smooth transition between them which may be an indication of the two different regimes of acceleration specially coexisting at larger energies, the reconnection and later drift acceleration regimes we identified in Figure \ref{pic1}. Interestingly, the power-law tail of the flatter  part of the spectrum  for $t= 45$ L/c, when  the $\sigma \sim 1$ jet  develops a fully turbulent regime, is very   similar to   the slope obtained in the  snapshot $t=50$ L/c in \citetalias{medinatorrejon_etal_2021} which is in a similar dynamical state  of the background jet (see their Figure 11). 
For the $\sigma \sim 10$ jet, the transition is more abrupt and characterized by large  humps around 6000 and 10000 $E_p/m_p c^2$. Examining the particles energy evolution in Figure \ref{pic1}, these humps seem to concentrate a substantial number of particles with acceleration of the parallel component predominantly, 
but the two regimes of acceleration also seem to coexist in these large energies, as indicated by the presence of  particles also with  the perpendicular component dominating the acceleration.  Clearly, for this model the amount of particles accelerated in this short dynamical time is comparatively smaller.
Since the acceleration of the particles is still in very early stages and far from reaching the saturation energy by reconnection, the large energy tails of these spectra are clearly still under development.   

\begin{figure} 
\centering
    \includegraphics[scale=0.7]{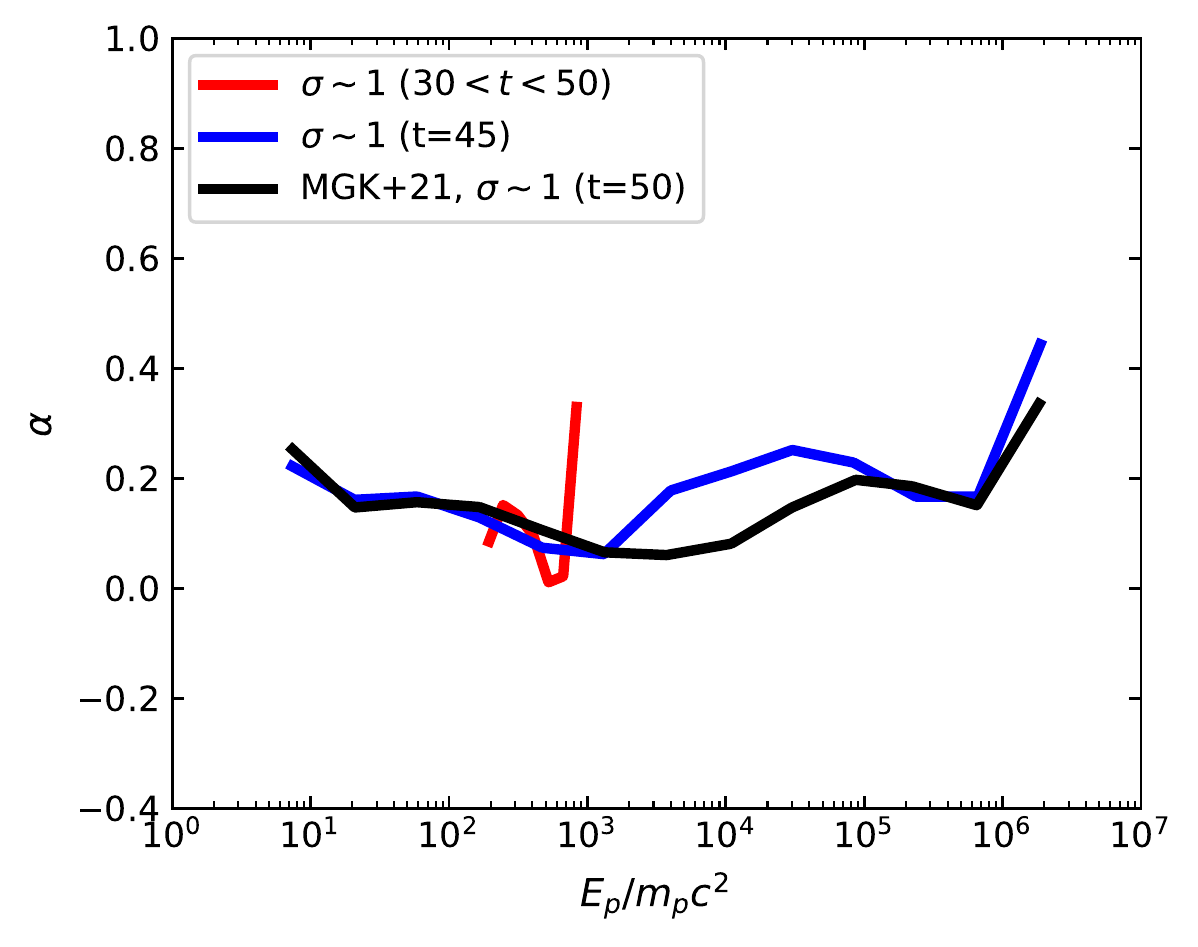}
\caption{Power-law index $\alpha = \Delta (\log t)/\Delta (\log E_p)$ of the acceleration time as function of the particle kinetic energy normalized by the proton rest mass energy. The minimum in the curves, $\alpha \sim 0.1$, indicates the nearly exponential regime of particle acceleration. Depicted are the models with steady-state turbulent background of Figure \ref{GACELL-PLUTO45}, namely  the  $\sigma \sim 1$ jet at  t=50 L/c run by \citetalias{medinatorrejon_etal_2021}  
(black line) and the $\sigma \sim 1$ jet  at  t = 45 L/c run in this work 
(blue line). Also shown is  $\alpha$  
for the nearly exponential regime  (between   $30 L/c<t<50 L/c$)
of the $\sigma \sim 1$ MHD-PIC model of the top of Figure \ref{pic1} where particles evolved with the background plasma (red curve).
}\label{alpha}
\end{figure}


Finally, we can quantify and compare the particle acceleration, in particular, in the nearly exponential regime, by evaluating the acceleration time directly from the diagrams of particles kinetic energy versus  time, in a similar way as performed previously
in \citet*{delvalle_etal_16} and \citetalias{medinatorrejon_etal_2021}. Specifically, we  compute
the slope of the logarithmic diagrams in Figures \ref{GACELL-PLUTO45} and \ref{pic1} (top), $\alpha = \Delta (\log t)/\Delta (\log E_p)$, which gives the acceleration time dependence with particle energy, $t_{acc} \propto E_p^ \alpha$.  The result is shown in Figure \ref{alpha}. We find that the slope $\alpha$ has essentially the same minimum value in all models, which corresponds to the nearly exponential regime of the acceleration of the particles, i.e., $\alpha \sim 0.1$, implying an acceleration time $t_{acc} \propto E_p^{0.1}$, as found in \citetalias{medinatorrejon_etal_2021}, with very weak dependence on the energy, as expected in this regime. The increase in $\alpha$ (and thus  in the acceleration time) around $E_p/m_pc^2 \sim 10^3$ for the MHD-PIC model is due to the contribution of several  particles that are already experiencing drift and thus slower acceleration at this energy (see the blue points in Figure \ref{pic1}  that correspond to the perpendicular  momentum component, predominant in drift acceleration).

~\\
\section{Discussion and Conclusions}\label{sec:discut}

In this work, we have investigated the early stages of the acceleration of the particles in  3D Poynting flux dominated jets with  magnetization $\sigma \sim$  1 and 10,  subject to CDKI, using the MHD-PIC mode of the \texttt{PLUTO} code, in order to follow the evolution of the particles along with the flow. The CDKI drives turbulence and fast magnetic reconnection which we find to be the dominant mechanism of particle acceleration. 

Our results are very similar to those of \citetalias{medinatorrejon_etal_2021} which were carried out with test particles launched in the simulated MHD relativistic jet after it achieved a 
regime of fully developed  turbulence.  Particles are accelerated by the ideal electric field ($V \times B$) of the background fluctuations, over the entire inertial range of the turbulence, starting in the small, resistive scales up to the large injection scales (Figure \ref{spectrum}). The connection of the accelerated particles with the magnetic reconnection layers is clear (Figure \ref{jet_points}). During this regime, the particles energy grow nearly exponentially and the parallel velocity component to the local magnetic field is the one that is preferentially accelerated, both expected in a Fermi-type process. In the test particle simulations of \citetalias{medinatorrejon_etal_2021} (see also Figure \ref{GACELL-PLUTO45}), particles re-enter the system several times through the periodic boundaries of the nearly steady state turbulent jet and are  accelerated in the reconnection sites up to the saturation energy that is achieved when their Larmor radius becomes of the order of the size of the acceleration region, or the jet diameter. This takes several hundred hours in the $\sigma \sim 1$ jet and the particles energy become as large $\sim 10^ 7$ $m_pc^2$. Beyond this energy, particles still experience further acceleration, but at smaller rate due to drift in the large scale non-reconnected fields.   In the MHD-PIC simulations, we can follow particle acceleration only during the dynamical time evolution of the MHD jet which  lasts $\sim 60$ L/c  and $\sim 35$ L/c for the $\sigma \sim 1$  and $\sigma \sim 10$ jet, respectively, and  corresponds to   only  $\sim 1 hr$ and half-hour, respectively, in physical units for the particles. During this time, the particles obviously do not reach the maximum possible (saturation) energy, 
but follow the same  exponential acceleration trend as in the test particle simulations
(Figure \ref{pic1}). 

At later times, when turbulence is fully developed, the particle energy  spectrum develops a power law tail with  two slopes (better defined in  the $\sigma \sim 1$ jet), suggesting the presence of the two different regimes of acceleration, the reconnection and the drift regimes (Figure \ref{picmhd-NE}).
The slope of the power-law tail of the flatter  part of the spectrum  for $t= 45$ L/c in the $\sigma \sim 1$  is the same as obtained for particles accelerating in the  snapshot $t=50$ L/c in \citetalias{medinatorrejon_etal_2021}, which has a similar state  of the background jet (see their Figure 11). These slopes are also comparable to  previous studies of particle acceleration both in MHD flows \citep{kowal_etal_2012,delvalle_etal_16} and PIC simulations \citep[e.g.,][]{comisso18, werner_etal_2018}. However, we expect that in realistic systems, the presence of 
radiative losses and dynamical feedback of the accelerated particles into the plasma will lead to steepening of the spectra \citepalias[e.g.,][]{medinatorrejon_etal_2021}. 

Our results also indicate that the time evolution of the background magnetic field ($\partial  B/ \partial  t$) does not influence the acceleration of the particles. They enter the exponential regime of  acceleration in the same dynamical times of the jet in which turbulence becomes fully developed ($\sim 30$ L/c for the $\sigma \sim 1$  jet, and $\sim 15$ L/c for the $\sigma \sim 10$, respectively;  Figure \ref{pic1}), in agreement with the results of the MHD simulations with test particles injected in the nearly steady state turbulent jet in \citetalias{medinatorrejon_etal_2021} (see also Figure \ref{GACELL-PLUTO45}). The particles also undergo curvature drift acceleration in the initial stage of the CDKI when the jet column starts to wiggle in similar dynamical time both in the test particle $+$ MHD and in the MHD-PIC simulations. The background magnetic field time evolution effect, also known  as betatron acceleration, has been found to affect particle acceleration in pure turbulent flows only by a factor two in the acceleration rate \citep[e.g.,][]{dalpino_kowal_15}. Therefore, while it can be substantial in  very early times when particles are still undergoing linear drift acceleration, it is negligible in the more advanced times when exponential acceleration takes over. 

The increase of the jet magnetization by a factor 10, speeds up the growth of the CDKI which attains saturation in nearly half of the time (see Figure \ref{energy}) and  particles are accelerated to energies about 10 times larger, as also expected from PIC simulations \citep[e.g.][]{werner_etal_2018}.

The  results  above indicate that particle acceleration by fast magnetic reconnection in a Fermi process can be dominant in magnetically dominated flows from the injection (large) to the resistive (small) scales of the turbulence. These results \citep[and those produced in earlier MHD works with test particles; e.g.][]{kowal_etal_2012,delvalle_etal_16,medinatorrejon_etal_2021} 
are in contrast with recent studies based on 3D PIC simulations that suggest that acceleration by reconnection would be  dominant only in the very early stages of particle energizing \citep[e.g.,][]{comisso19,sironi_etal_2021,sironi2022,comisso_sironi_2022}. 
This apparent inconsistency is essentially due to the intrinsic difference in scales and in the accelerating electric fields that prevail in the two regimes. 
While in these PIC simulations, plasmoid-like reconnection acceleration occurs at the small kinetic, resistive scales and is dominated by the resistive electric field ($\eta J$, where $\eta$ is the resistivity  and $J$  the current density), in our collisional MHD turbulent flow simulations where  resistivity is naturally small (the ubiquitous Ohmic resistivity is mimicked by the numerical truncation error), the reconnection layers persist up to the large injection scales and particles are accelerated by the ideal electric fields (V$\times B$) of the fluctuations in these sites.  
Therefore, these intrinsic differences (inherent to scale and accelerating electric field),    indicate that direct extrapolation from
the resistive small scales probed by PIC simulations (wherein non-ideal accelerating electric fields
generally prevail), to the large MHD scales should be taken with caution \citep[see also][]{guo_etal_2019, guo_etal_2022}.

The same applies to the recent study of \citet*{puzzoni_etal_2022} who examined the impact of resistive electric fields on particle acceleration in reconnection layers. The authors claimed  that their results are  in contradiction with  earlier MHD works \citep*{kowal_etal_2011,kowal_etal_2012,medinatorrejon_etal_2021}. However, they are clearly exploring a different regime of reconnection endowed with extremely high artificial resistivity, which is much larger than the Ohmic resistivity expected in most astrophysical  MHD flows and in particular, in turbulent ones. In other words, they are exploring the resisistive, kinetic  scales well below the inertial range of the turbulence that is explored in the works above and in the present one. 
While in the present simulations and those of the previous works mentioned above, particles are predominantly accelerated by the ideal electric fields of the magnetic fluctuations in the reconnection layers, in \citep{puzzoni_etal_2022} simulations, the dominant component is the resistive electric field component 
which prevails in the kinetic scales. Therefore, \textbf{there is no contradiction} with the MHD (non-resistive) works above.\footnote{One may still inquire how the results of the present study would change if we had included an explicit resistivity in the flow.  As remarked above, this would affect only the very small scales of the flow, of the order of a few grid cells size \citep[e.g.][]{santoslima_etal_2010}. In the integration of the particles equation of motion, we accounted only for the ideal electric fields of the magnetic fluctuations that persist in the entire range of the turbulence.  Still, the non-ideal term could be important for the small-scale topology of the velocity and magnetic fields, especially in the vicinity of the reconnection regions, indirectly affecting the particles' evolution before they reach a gyroradius of the order of a few cells size.
Therefore, if we had included an initial small explicit resistivity of the typical strength of Ohmic resistivity (as expected in astrophysical turbulent flows), the results for particle acceleration would be the same as in the present work.  On the other hand, if we had adopted an artificial much larger explicit resistivity, well above the Ohmic resistivity, this would kill all the turbulence in the range of scales smaller than this resistive scale and particle acceleration by turbulent reconnection would be possible only in a more limited inertial range of turbulent structures, from the injection scale down to the resistive scale.}

Future studies exploring in depth both regimes and scales, and also including particle feedback into the plasma are required. Our present study, combining PIC and MHD altogether in a relativistic jet with turbulence induced by an instability was a first attempt in this direction and the results in general confirm the predictions of previous MHD studies with test particles which show that  turbulent reconnection acceleration prevails in most of the scales of the system.  As stressed, e.g. in \citetalias{medinatorrejon_etal_2021}, the implications of these results for particle acceleration and the origin of VHE emission phenomena  in  Poynting flux dominated systems like the relativistic jets in microquasars, AGN and GRBs, is rather important.



\acknowledgments
 The authors acknowledge very useful discussions with L. Kadowaki.  TEMT and EMdGDP acknowledge support   from the Brazilian Funding Agency FAPESP (grant 13/10559-5),  EMdGDP also acknowledges support  from CNPq (grant 308643/2017-8),  and G.K. from  FAPESP (grants 2013/10559-5, 2019/03301-8, and 2021/06502-4).  The  simulations presented in this work were performed in the cluster of the Group of Plasmas and High-Energy Astrophysics (GAPAE), acquired with support from  FAPESP (grant 2013/10559-5), 
 and  the computing facilities of the Laboratory of Astroinformatics (IAG/USP, NAT/Unicsul), whose purchase was also made possible by FAPESP (grant 2009/54006-4) and the INCT-A. 

\bibliography{bibliography.bib}



\end{document}